\documentclass[nofootinbib,pra,twocolumn,showpacs,superscriptaddress,notitlepage,amsmath,amssymb]{revtex4-1}
\usepackage{amsmath,amssymb}
\usepackage{graphicx}
\usepackage[colorlinks=true,linkcolor=blue,citecolor=blue,urlcolor=blue]{hyperref}
\usepackage{xcolor}
\usepackage{pifont}
\usepackage{braket}
\usepackage{bm}
\usepackage{algorithm}
\usepackage{enumitem}

\usepackage[normalem]{ulem}

\usepackage[colorlinks=true,citecolor=blue,linkcolor=blue,urlcolor=blue]{hyperref}

\usepackage{framed}
\definecolor{shadecolor}{RGB}{224,224,224}

\definecolor{DarkGreen}{RGB}{0,128,0}

\usepackage{tikz}
\usepackage{pgfplots}
\usetikzlibrary{fadings}
\usetikzlibrary{plotmarks}

\usetikzlibrary{shapes.geometric}

\begin{document}

\title{Integer Linear Programming Decoder for Abelian and Non-Abelian Topological Codes}

\author{Dian Jing}
\email{rossoneri@uchicago.edu}
\affiliation{Department of Physics, University of Chicago, Chicago, IL 60637, USA}
\affiliation{Pritzker School of Molecular Engineering, University of Chicago, Chicago, IL 60637, USA}

\author{Aubrey Zhang}
\affiliation{Department of Physics, University of Chicago, Chicago, IL 60637, USA}

\author{Liang Jiang}
\affiliation{Pritzker School of Molecular Engineering, University of Chicago, Chicago, IL 60637, USA}

\author{Ruben Verresen}
\affiliation{Pritzker School of Molecular Engineering, University of Chicago, Chicago, IL 60637, USA}

\date{\today}

\begin{abstract}
Topological orders (TOs) are widely used as quantum error-correcting codes, with anyon excitations serving as error syndromes. For certain Abelian TOs, decoding can be performed by independently matching particle-antiparticle pairs of each species. However, matching-based decoders cannot handle more general fusion rules in either Abelian or non-Abelian TOs, nor account for noise that correlates different anyon species. While clustering decoders are more broadly applicable, they typically neglect anyon data and fusion properties, leading to poor performance in practice. In this work, we introduce a fundamentally different decoder for arbitrary TOs based on integer linear programming (ILP). The ILP formulation linearizes the error-correction problem through the introduction of auxiliary variables and encodes fusion rules as linear constraints. Classical optimization then identifies the minimum-weight error configuration. As concrete examples, we determine error-correction thresholds for three TOs: the Abelian $\mathbb{Z}_2$ TO under depolarizing noise, where charge and flux errors are correlated; the Abelian $\mathbb{Z}_3$ TO, which does not admit a pairwise matching decoder; and the non-Abelian $D_4$ TO under noise channels that generate all anyon species. We demonstrate the versatility of the ILP decoder by showing a clear performance advantage over most existing decoders in all three cases. We further extend the method to incorporate noisy syndrome measurements and propose a just-in-time variant for continuous error correction. Our results establish ILP as a natural framework for handling correlated errors and general anyon fusion rules, and as a powerful and flexible general-purpose decoder for incoherent anyon noise in arbitrary TOs, with applications to fault-tolerant quantum computation.
\end{abstract}

\maketitle

\section{Introduction}
\begin{figure}[t]
	\includegraphics[width=1\columnwidth]{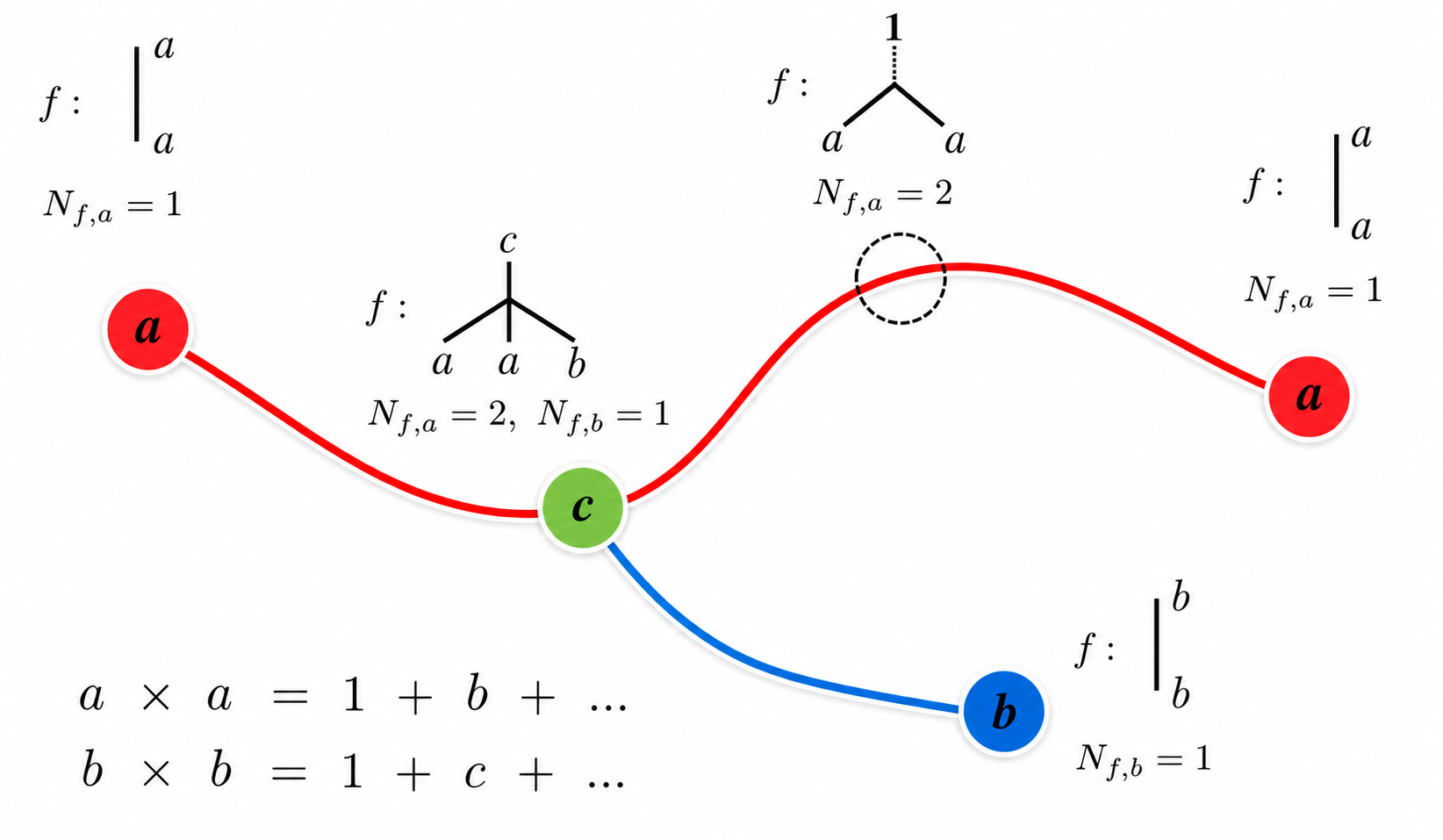}
	\caption{\textbf{Example configuration of incoherent anyon errors obtained from the ILP decoder.} Given the anyon syndromes, the ILP decoder uses classical optimization to identify the minimum-weight configuration of incoherent anyon errors. These configurations allow anyon strings to terminate on syndromes as well as other anyon strings according to the fusion rules of the TO, a feature not captured by existing decoders. In the ILP formulation, the fusion channels $f$ and the corresponding multiplicities $N_f$ of incident anyons are encoded as linear constraints. An example anyon-string configuration is shown, with the relevant fusion rules provided in the bottom-left panel. At each lattice site with a measured anyon syndrome, the selected fusion channel $f$ and the multiplicities $N_f$ of incident anyons are specified. The labels at the bottom of each fusion channel $f$ denote the incident anyons, while the label at the top denotes the fusion outcome corresponding to the measured syndrome.}
	\label{Example}
\end{figure}

\indent Topological orders (TOs) are long-range entangled quantum phases that are robust against local perturbations \cite{Leinaas1977, Goldin81, Wilczek82, Goldin85, Moore1989, Wen90, Wen91, MOORE1991362, Einarsson95}. They have been exploited as quantum error correction codes, with their anyon excitations serving as error syndromes and anyon strings acting as correction operators \cite{KITAEV20032, Freedman2002, Freedman06, RevModPhys2008, Dennis, RMP2015}. TOs are either Abelian or non-Abelian. For Abelian TOs, such as the toric code, anyons obey deterministic fusion rules and pairwise annihilate with their antiparticles. Their error correction problem is well studied, and a variety of efficient decoding algorithms have been developed \cite{Dennis, Raussendorf, RMP2015, BravyiHaah, HarringtonThesis}.\\
\indent Decoders for Abelian TOs can be broadly divided into two categories: matching-based decoders and clustering decoders. For anyons that are their own antiparticles, matching-based decoders identify lattice paths along which they can be pairwise annihilated. Classical graph algorithms, such as Dijkstra's algorithm \cite{DIJKSTRA1959, higgott2021pymatchingpythonpackagedecoding} or the Blossom algorithm \cite{BlossomV, BurtonIsing}, are typically used to determine the shortest pairing paths, yielding the strategy known as minimum-weight perfect matching (MWPM) \cite{Dennis, WANG200331}. However, matching-based decoders apply only in special cases: they cannot naturally accommodate more general fusion rules, even for Abelian TOs in which an anyon is not its own antiparticle, and become inadequate when different anyon species are correlated by the noise. In contrast, clustering decoders identify and fuse anyons within independent error clusters created by errors with non-overlapping support \cite{GACS198615, Gacs89, Gacs2001}. While clustering decoders apply to arbitrary TOs and arbitrary local noise channels, their under-utilization of anyon data, fusion properties, and details of the noise channel typically leads to lower thresholds in practice \cite{HarringtonThesis, BravyiHaah}. Since the anyons within each cluster necessarily fuse to the vacuum, independent of their internal braiding or fusion structure, logical errors are avoided provided that all clusters remain smaller than the code distance. Consequently, clustering decoders exhibit subthreshold behavior whenever the probability of large error clusters is sufficiently suppressed \cite{HarringtonThesis, Wootton2016proof, Dauphinais2017}.\\
\indent While decoding general Abelian TOs is already challenging, error correction for non-Abelian TOs is even more difficult due to their non-Abelian braiding and fusion properties \cite{Pachos_2012}. By definition, non-Abelian anyons, denoted by $a$, have multiple, nondeterministic fusion channels
\begin{equation}
	a \; \times \; \Bar{a} \; = \; 1 \; + (a) \; + \; b \; + \; ... \; ,
	\label{fusion}
\end{equation}
leading to a quantum dimension $d_a > 1$. Moreover, an anyon $a$ may appear among its own fusion outcomes (as indicated by the parentheses), a feature commonly referred to as the cyclic property.\\
\indent For non-Abelian TOs, matching-based decoders are applicable only to anyons that are their own antiparticles and obey acyclic fusion rules \cite{BurtonIsing, davydova2026, jing2025intrinsicheralding, delafuente2026}. In this restricted setting, intrinsic heralding can provide additional information about error paths and thereby improve error-correction thresholds by exploiting nondeterministic fusion outcomes \cite{jing2025intrinsicheralding}. However, intrinsic heralding is effective only when the noise channel is dominated by pair creation of a single species of acyclic, self-antiparticle non-Abelian anyon. More severely, for cyclic non-Abelian TOs, which can support universal quantum computation through braiding and fusion \cite{Mochon04, Lo25, Chen2025, Minev2025, lo2026universalgatesbraidingfusing, RevModPhys2008, Freedman2002}, matching-based decoding breaks down entirely, even when the anyons are their own antiparticles.\\
\indent In general, non-Abelian anyon strings can form branching, tree-like networks that cannot be decomposed into pairwise annihilation. As a result, existing decoders for non-Abelian TOs are almost exclusively clustering decoders \cite{WoottonLoss, Wootton_improved_HDRG, BurtonIsing, BurtonFibonacci, Verstraete, Dauphinais2017, lyons2026}, in which independent error clusters are identified using renormalization group (RG) methods \cite{WoottonSimple, HarringtonThesis}. While such clustering decoders have provided valuable tools for proving the existence of error correction thresholds for non-Abelian TOs \cite{HarringtonThesis, Wootton2016proof, Dauphinais2017, davydova2026, lyons2026} and for numerically demonstrating threshold behavior, they do not utilize the fusion and braiding properties of the underlying TO. Consequently, their numerical thresholds are typically much lower than the MWPM thresholds achieved for Abelian TOs \cite{HarringtonThesis, Dauphinais2017, HarringtonFibonacci, WoottonLoss, Wootton_improved_HDRG, BurtonIsing, BurtonFibonacci, Verstraete}.\\
\indent These limitations motivate the development of fundamentally new decoding strategies applicable to both Abelian and non-Abelian anyons. To this end, we introduce a qualitatively different decoding approach based on the classical optimization framework of integer linear programming (ILP). To date, linear programming decoders have been primarily explored in the context of qLDPC codes \cite{Feldman2003UsingLP, Feldman2005UsingLP, Li2018LP, Fawzi2021LP, javed2024lowcomplexitylinearprogrammingbased, gu2025linearprogramming}, whose syndromes effectively behave as Abelian anyons with the simple fusion rule of particle-antiparticle annihilation. In this setting, the expressive power of ILP is not fully utilized, and decoding is more efficiently and preferably handled by other methods such as belief propagation \cite{classicalBP, quantumBP1, quantumBP2}. By contrast, for arbitrary TOs with general fusion rules, ILP provides a genuine extension beyond pairwise annihilation of anyon syndromes. It does so by explicitly encoding fusion rules as linear constraints on anyon-string configurations, thereby capturing allowed string terminations that are absent from existing decoding schemes, as illustrated in Fig.~\ref{Example}. These linear constraints can also encode correlations among different anyon species in the local noise channel. Although auxiliary variables are required to represent the fusion rules and correlations, the ILP formulation linearizes the error-correction problem, enabling classical optimization to identify the most probable error configuration consistent with the measured syndromes. This yields a minimum-weight decoder for arbitrary TOs that serves as a direct analog of MWPM. In this way, we establish linear constraints as a natural language for enforcing anyon fusion rules and ILP as a natural framework for general-purpose decoding of arbitrary TOs.\\
\indent While our construction is general, we benchmark the performance of the ILP decoder by numerically comparing its error-correction threshold against that of existing decoders for the Abelian $\mathbb{Z}_2$ TO, the Abelian $\mathbb{Z}_3$ TO, and the acyclic non-Abelian $D_4 \cong \mathbb{Z}_4 \rtimes \mathbb{Z}_2$ TO. As a warm-up and an introduction to the ILP formalism, we first consider decoding the $\mathbb{Z}_2$ TO under a depolarizing channel that correlates charge and flux errors. We then consider the Abelian $\mathbb{Z}_3$ TO, whose anyon fusion rules do not allow a matching-based decoder, under an incoherent noise channel in the charge sector. Finally, we consider the non-Abelian $D_4$ TO under a noise channel that independently and incoherently creates all anyon species. The $D_4$ TO is of practical relevance, having recently been realized in trapped-ion experiments \cite{D4preparation, Iqbal2024}, and provides a resource for universal quantum computation \cite{davydova2026, D4nonCliffordOxford, D4nonCliffordEllison}. Importantly, the self-antiparticle and acyclicity properties of the $D_4$ fusion rules permit matching-based decoding, enabling a direct comparison. Since clustering decoders are known to exhibit substantially lower thresholds than matching-based decoders, demonstrating an advantage over matching suffices to establish the performance gains of the ILP approach.
\section{Minimum-weight Decoding with Integer Linear Programming}
\indent Given a set of anyon measurement outcomes $\bm{\sigma}$ serving as error syndromes, the goal of error correction is to identify the most likely homology class $h$ of physical error strings across all anyon species \cite{Dennis}, according to
\begin{align}
    P(h|\bm{\sigma}) = & \sum_{E\in h} P(E|{\bm{\sigma}}) \propto \sum_{E\in h} P(\bm{\sigma}|E)P(E) \nonumber\\
    \approx & \max_{E\in h} P(\bm{\sigma}|E)P(E).
    \label{homology}
\end{align}
In the second step, the probability of a physical error string $E$ conditioned on the syndrome $\bm{\sigma}$ is decomposed via Bayes' rule into the prior $P(E)$ and the likelihood $P(\bm{\sigma}|E)$, where the latter captures the probabilistic collapse of any superposition of non-Abelian fusion outcomes into the observed syndromes $\bm{\sigma}$ \cite{jing2025intrinsicheralding}. For Abelian TOs, the fusion outcomes are deterministic, so $P(\bm{\sigma}|E)$ reduces to a delta function that enforces consistency between the error configuration $E$ and the measured syndrome $\bm{\sigma}$. Approximating the sum over all error strings $E$ by its dominant contribution, as in the final step of Eq.~\ref{homology}, leads to the minimum-weight decoder.\\
\indent For Abelian TOs with the simple fusion rule of particle-antiparticle annihilation, decoding can be implemented efficiently using MWPM \cite{WANG200331}, which, in the context of this paper, specifically refers to a particle-antiparticle matching decoder restricted to a single anyon species. In contrast, minimum-weight decoding for arbitrary TOs, either Abelian or non-Abelian, is more challenging due to general anyon fusion rules, which allow error strings to terminate on different anyon species. This difficulty is especially pronounced for non-Abelian TOs, where nondeterministic fusion channels give rise to a nontrivial $P(\bm{\sigma}|E)$. In this work, we develop an ILP approach for minimum-weight decoding of arbitrary TOs under incoherent noise that can generate all anyon species, including correlated errors across different species, thereby naturally addressing both of these difficulties.\\
\indent The linear programming problem \cite{ILPtextbook} seeks to optimize a linear objective function
\begin{equation}
	f(\bm{x})=\bm{w}^{\mathsf T} \bm{x}=w_1x_1 + w_2x_2 + \ldots,
	\label{objective}
\end{equation}
where $\bm{w}$ is a vector of real coefficients and $\bm{x}$ is the vector of decision variables subject to the linear constraints
\begin{equation}
	\bm{b}_L \leq \bm{Ax} \leq \bm{b}_R, \;\;\;\; \bm{x} \geq \bm{0}.
	\label{constraint}
\end{equation}
If the decision variables are restricted to integer or binary values, the problem is referred to as integer linear programming (ILP) or binary linear programming (BLP), respectively.\\
\indent The decision variables $\bm{x}$ will encode both the physical error-string configuration $E$ and the anyon fusion outcomes along the strings. The linear constraints, specified by $\bm{A}$, $\bm{b}_L$, and $\bm{b}_R$ in Eq.~\ref{constraint}, ensure that the decision variables are consistent with the fusion rules of the TO and the measured syndromes $\bm{\sigma}$. Thus, the error configuration is determined by the decision variables, $E=E(\bm{x})$, while the constraint bounds depend on $\bm{\sigma}$. Moreover, the constructions of $\bm{x}$, $\bm{A}$, $\bm{b}_L$, and $\bm{b}_R$ depend on the anyon data and fusion rules of the TO.\\
\indent For minimum-weight error correction in TOs, the coefficients $\bm{w}$, also referred to as weights, are chosen so that the objective function $f(\bm{x})$ is a monotonic function of $P(\bm{\sigma}|E)P(E)$ in Eq.~\ref{homology}. Under local noise channels, the probability $P(E|\bm{\sigma}) \propto P(\bm{\sigma}|E)P(E)$ of an error configuration is built from products of local probabilities, while the objective function is additive. This motivates choosing $f(\bm{x})=\ln \left[P(\bm{\sigma}|E)P(E)\right]$. Therefore, optimizing $f(\bm{x})$ correctly identifies $\max_E P(\bm{\sigma}|E)P(E)$.\\
\indent Details of the ILP decoder, including its decision variables, linear constraints, and objective functions, are presented in Secs.~\ref{Sec:Z2}-\ref{Sec:D4} for general TOs with perfect syndrome measurements, together with explicit examples for the Abelian $\mathbb{Z}_2$ and $\mathbb{Z}_3$ TOs and the non-Abelian $D_4$ TO. Continuous spacetime ILP decoding in the presence of measurement errors is discussed in Sec.~\ref{measure_error}.
\section{ILP for Abelian $\mathbb{Z}_2$ TO under correlated noise}
\label{Sec:Z2}
\indent As a warm-up, we introduce a simplified ILP formalism using the Abelian $\mathbb{Z}_2$ TO, commonly known as the toric code, under single-qubit depolarizing noise with equal Pauli probabilities, $p_X=p_Y=p_Z=\frac{p}{3}$. While Pauli $\hat{X}$ and $\hat{Z}$ operators incoherently create $m$- and $e$-anyons, respectively, Pauli $\hat{Y}$ operators create both species of anyons in a correlated manner. Therefore, the $\mathbb{Z}_2$ example also serves as a demonstration of the ability of ILP to correct correlated anyon errors.
\subsection{Input}
\indent We consider the $\mathbb{Z}_2$ TO on a square lattice with periodic boundary conditions, with qubits residing on lattice links $\ell$. As in Ref.~\onlinecite{KITAEV20032}, a site $s$ is defined by a vertex together with the plaquette immediately northeast of that vertex. We denote by $\partial_e s$ the set of qubits in the support of the vertex operator, on which a Pauli $\hat{Y}$ or $\hat{Z}$ operator creates an $e$-anyon at site $s$, and by $\partial_m s$ the set of qubits in the support of the plaquette operator, on which a Pauli $\hat{X}$ or $\hat{Y}$ operator creates an $m$-anyon at site $s$.\\
\indent At each site, perfect anyon measurement is performed, reporting a syndrome of either vacuum, $e$, $m$, or $f$. Although this measurement scheme is equivalent to independently measuring $e$-anyons at all vertices and $m$-anyons at all plaquettes, this representation includes syndrome labels corresponding to fusion channels beyond simple particle-antiparticle annihilation, namely $f=e\times m$, making it an illustrative minimal example of ILP decoding.\\
\indent For each decoder instance, the syndromes are input to the decoder as binary variables $\sigma^e_s,\sigma^m_s,\sigma^f_s\in\{0,1\}$, where $\sigma^{e/m/f}_s=1$ indicates the presence of an $e/m/f$-anyon syndrome and $\sigma^{e/m/f}_s=0$ indicates its absence.
\subsection{Weights and Objective Function}
\indent To describe errors from the depolarizing noise channel, we introduce three binary error variables $y^X_{\ell}, y^Y_{\ell}, y^Z_{\ell}\in\{0,1\}$ for each qubit $\ell$. The assignment $y^P_{\ell}=1$ indicates that the Pauli operator $P\in\{X,Y,Z\}$ acts on qubit $\ell$, while the all-zero assignment corresponds to the identity.\\
\indent Under depolarizing noise, each of the three nontrivial Pauli errors occurs with probability $p/3$, while the identity occurs with probability $1-p$. Therefore, an error configuration $E$, uniquely specified by the assignment of the binary error variables $\bm{y}$, occurs with probability
\begin{equation}
    P(E)=\prod_\ell (1-p) \left(\frac{p/3}{1-p}\right)^{y^X_{\ell}} \left(\frac{p/3}{1-p}\right)^{y^Y_{\ell}} \left(\frac{p/3}{1-p}\right)^{y^Z_{\ell}}.
\end{equation}
\indent For the Abelian $\mathbb{Z}_2$ TO, the deterministic fusion rules imply that $P(\bm{\sigma}|E)$ is either zero or one, depending on whether $E$ is consistent with the syndrome $\bm{\sigma}$. The constraints introduced in the next subsection enforce this consistency, ensuring that every candidate error configuration $E$ satisfies $P(\bm{\sigma}|E)=1$. It is therefore sufficient to optimize $P(E)$ and assign weights only to the error variables $\bm{y}$.\\
\indent Taking the logarithm $f(\bm{y})=\ln \left[P(\bm{\sigma}|E)P(E)\right]$ and restricting to allowed configurations satisfying $P(\bm{\sigma}|E)=1$, we obtain, up to an additive constant, $f(\bm{y})=\sum_{\ell,P} w^P_{\ell}y^P_{\ell}$, where $w^P_{\ell} = \ln \frac{p/3}{1 - p} <0$ is the weight associated with an activated error variable $y^P_{\ell}=1$. Since this weight is identical for all error variables, we may rescale the objective function by the common negative weight, reverse the optimization direction, and use the equivalent objective function
\begin{equation}
    f(\bm{y})=\sum_{\ell}\left(
	y^X_{\ell}+y^Y_{\ell}+y^Z_{\ell}
	\right).
    \label{Z2_objective}
\end{equation}
Minimum-weight decoding is achieved by minimizing $f(\bm{y})$, yielding an assignment of $\bm{y}$ that corresponds to $\max_E P(\bm{\sigma}|E)P(E)$. Intuitively, the ILP decoder identifies the most likely error configuration consistent with the measured syndrome, which in this case is the configuration containing the fewest Pauli errors.
\subsection{Linear Constraints}
\indent The core of the ILP decoder lies in the decision variables and linear constraints in Eq.~\ref{constraint}, which ensure that each allowed assignment of the decision variables encodes an anyon-string configuration consistent with both the measured syndromes and the fusion rules of the underlying TO.\\
\indent For the $\mathbb{Z}_2$ TO, the anyon fusion rules impose parity constraints on the Pauli errors incident on each site $s$. On the support $\partial_e s$ of the vertex, the total number of Pauli $\hat{Y}$ and $\hat{Z}$ operators must be odd when the measured syndrome is $e$ or $f$, and even when it is $m$ or vacuum. Similarly, on the support $\partial_m s$ of the plaquette, the total number of Pauli $\hat{X}$ and $\hat{Y}$ operators must be odd when the measured syndrome is $m$ or $f$, and even otherwise. These fusion rules are encoded by the constraints
\begin{align}
\sum_{\ell\in\partial_e s} \left(y^Y_{\ell}+y^Z_{\ell}\right)
&=\sigma^e_s+\sigma^f_s+2n^e_s, \nonumber\\
\sum_{\ell\in\partial_m s} \left(y^X_{\ell}+y^Y_{\ell}\right)
&=\sigma^m_s+\sigma^f_s+2n^m_s,
\label{tc_constraint}
\end{align}
defined at each site $s$. The integer variables $n^e_s,n^m_s\in\{0,1,2\}$ enforce the required parities. They are necessary because the matrix multiplication $\bm A\bm x$ in Eq.~\ref{constraint} cannot directly implement the modulo-$2$ addition needed to impose parity constraints. The bounds on $n^e_s$ and $n^m_s$ are sufficient because each vertex and plaquette operator is supported on four qubits.\\
\indent Since the single-qubit Pauli noise channel applies at most one nontrivial Pauli operator to each qubit, one may impose the constraint
\begin{equation}
y^X_{\ell}+y^Y_{\ell}+y^Z_{\ell}\leq 1
\label{tc_single_pauli}
\end{equation}
for every qubit $\ell$. However, this constraint is redundant under minimization of the objective function in Eq.~\ref{Z2_objective}. Suppose an ILP solution satisfying Eq.~\ref{tc_constraint} has two active error variables on the same qubit, for example $y^X_\ell=y^Y_\ell=1$ and $y^Z_\ell=0$. Since the product of Pauli $\hat{X}$ and $\hat{Y}$ operators is equivalent, up to a global phase, to a Pauli $\hat{Z}$ operator, the assignment $y^X_\ell=y^Y_\ell=0$ and $y^Z_\ell=1$ produces the same syndrome and also satisfies Eq.~\ref{tc_constraint}, with an appropriate reassignment of $n^e_s$ and $n^m_s$, while reducing the objective $f(\bm{y})$. The ILP decoder therefore always prefers the latter assignment. Similarly, the assignment $y^X_\ell=y^Y_\ell=y^Z_\ell=0$ is always preferred over $y^X_\ell=y^Y_\ell=y^Z_\ell=1$. This automatic satisfaction of Eq.~\ref{tc_single_pauli} is specific to the $\mathbb{Z}_2$ TO under single-qubit Pauli noise. We nevertheless include the constraint here as an example of a linear constraint on the error variables imposed directly by the noise model.

\subsection{Decoder Performance}
\indent In summary, the ILP decoder for the $\mathbb{Z}_2$ TO under the depolarizing noise channel minimizes the objective function $f(\bm{y})$ in Eq.~\ref{Z2_objective}, which depends on the error variables $\bm{y}$ contained in the decision-variable vector $\bm{x}^T=\left(\bm{y}_\ell^P, \; \bm{n}_s^e, \; \bm{n}_s^m\right)$. These decision variables are subject to linear constraints of the form in Eq.~\ref{constraint}, specifically Eq.~\ref{tc_constraint}, together with bounds on their allowed ranges.
\begin{figure}[t]
	\includegraphics[width=\columnwidth]{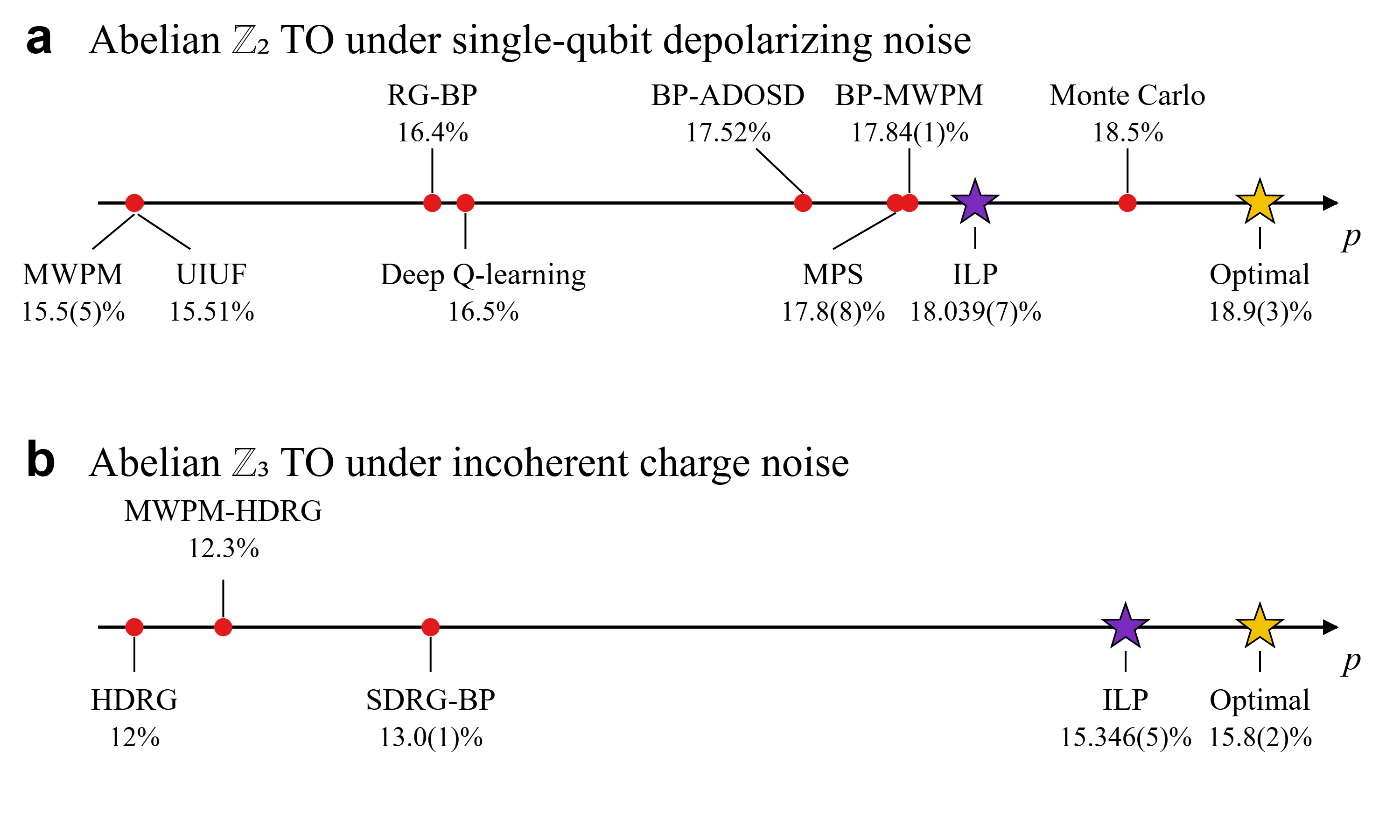}
	\caption{\textbf{Error-correction thresholds for example Abelian TOs.} (a) For the Abelian $\mathbb{Z}_2$ TO under single-qubit depolarizing noise, the ILP decoder achieves an error-correction threshold of $18.039(7)\%$, exceeding most existing decoders, including uncorrelated MWPM \cite{Z2-MWPM-20104}, union-intersection union-find (UIUF) \cite{Z2-UIUF-2025}, renormalization group with belief propagation (RG-BP) \cite{Z2-RG-BP-20102}, deep Q-learning based on deep reinforcement learning \cite{Z2-deepQ-2020}, belief propagation with approximate degenerate ordered-statistics decoding (BP-ADOSD) \cite{Z2-BPADOSD-2022, Z2-BPADOSD-2026}, matrix product states (MPS) \cite{Z2-MPS-2014}, and belief propagation with MWPM (BP-MWPM) \cite{Z2-BPMWPM-2018}. This threshold is also close to the optimal value of $18.9(3)\%$ \cite{BombinTC, OhzekiTC} and to that obtained using an approximate optimal decoder based on Monte Carlos sampling \cite{Z2-MC-2012}. (b) For the Abelian $\mathbb{Z}_3$ TO under incoherent charge noise, the ILP decoder achieves an error-correction threshold of $15.346(5)\%$, substantially exceeding those reported for existing decoders, including hard-decision renormalization group (HDRG) \cite{Z3_Anwar_2014}, MWPM-based HDRG \cite{Z3_Hutter_2015}, and soft-decision renormalization group (SDRG) \cite{Z3_Poulin_2013}. This threshold is also close to the optimal value of $15.8(2)\%$ estimated using Monte Carlo sampling \cite{Z3_optimal, Z3-Simon-2026}.}
	\label{Fig:Z2Z3}
\end{figure}

\indent With the error-correction problem formulated as an ILP, the minimum-weight error configuration can be obtained using commercial solvers such as Gurobi \cite{gurobi}. Numerically, this ILP decoder achieves a threshold of $18.039(7)\%$, outperforming most existing decoders \cite{Z2-RG-BP-20102, Z2-MPS-2014, Z2-BPMWPM-2018, Z2-deepQ-2020, Z2-BPADOSD-2022, Z2-UIUF-2025, Z2-BPADOSD-2026} and substantially exceeding the uncorrelated MWPM value of $15.5(5)\%$ \cite{Z2-MWPM-20104}, as shown in Fig.~\ref{Fig:Z2Z3}. This result is also close to the optimal depolarizing-noise threshold of $18.9(3)\%$ \cite{BombinTC, OhzekiTC}. Exact optimal decoding would require accounting for all possible error configurations, whose number grows exponentially with the code distance. Such an exhaustive calculation is infeasible in practice. Approximate optimal decoders therefore use Monte Carlo sampling to estimate the most likely error homology class \cite{Z2-MC-2012, BombinTC}. Although runtime depends on the specific code and implementation, the ILP decoder typically runs orders of magnitude faster than the approximately optimal Monte Carlo decoder at numerically relevant code distances while retaining near-optimal performance. Furthermore, it avoids the complicated convergence criteria associated with Monte Carlo sampling.\\
\indent The minimum-weight decoding problem for the $\mathbb{Z}_2$ TO under single-qubit Pauli noise is NP-hard \cite{NPhard1, NPhard2, NPcomplete}, and we efficiently formulate it as an ILP, which is itself NP-hard in general. Accordingly, polynomial-time scaling cannot be guaranteed in the worst case. Nevertheless, worst-case complexity does not determine the runtime for practically relevant error configurations. In practice, quantum error correction operates in the subthreshold regime, where we find that both the mean and median runtimes of our ILP decoder are consistent with polynomial scaling over the numerically accessible code distances. At the error-correction threshold, the runtime increases substantially, but the observed scaling remains inconsistent with a simple exponential dependence on system size, measured by the number of qubits, over the accessible range. Details of the runtime scaling are provided in Appendix~\ref{App:Z2}. These results indicate that the ILP decoder can operate efficiently over practically relevant error rates and code distances.
\section{ILP for Abelian $\mathbb{Z}_3$ TO with non-trivial fusion}
\label{Sec:Z3}
\indent Following the Abelian $\mathbb{Z}_2$ example, we now introduce a simplified ILP formulation for the Abelian $\mathbb{Z}_3$ TO. For simplicity, we focus on the charge sector, whose nontrivial anyons $e$ and $\bar e$ obey the deterministic fusion rules
\begin{equation}
	e \times \bar{e} = 1, \;\;\; e \times e = \bar{e},\;\;\; \bar{e} \times \bar{e} = e.
    \label{fusion_Z3}
\end{equation}
In particular, in addition to particle-antiparticle annihilation, three identical nontrivial anyons can fuse to the vacuum, e.g., $e\times e\times e=1$. Consequently, the $\mathbb{Z}_3$ TO cannot in general be decoded using a matching-based decoder. This example therefore demonstrates the ability of the ILP decoder to accommodate more general fusion rules while yielding a substantial threshold improvement over existing clustering decoders. We consider decoding the $\mathbb{Z}_3$ TO under incoherent creation of $e\bar e$ pairs on neighboring lattice sites, with the two possible pair orientations each occurring with probability $\frac{p}{2}$.

\subsection{Input}
\indent We consider the $\mathbb{Z}_3$ TO on a square lattice with periodic boundary conditions, with qutrits residing on lattice links $\ell$. Since the noise channel acting on the qutrits can create $e\bar e$ pairs in two orientations, we assign a fixed orientation to each qutrit and label the two error orientations by $e$ and $\bar e$. At a lattice site $s$, we denote by $\partial_+ s$ the set of qutrits oriented toward $s$, for which an error labeled by $e$ creates an $e$-anyon at site $s$, and by $\partial_- s$ the set of qutrits oriented away from $s$, for which an error labeled by $e$ creates a $\bar e$-anyon at site $s$. The choice of qutrit orientation is arbitrary provided that it is made consistently.\\
\indent At each site $s$, a perfect anyon measurement assigns a unique anyon charge label, including the vacuum. The corresponding syndrome data are supplied to the decoder as binary variables $\sigma^e_s,\sigma^{\bar e}_s\in\{0,1\}$, with $\sigma^e_s=1$ or $\sigma^{\bar e}_s=1$ indicating that an $e$-anyon or $\bar e$-anyon, respectively, is measured at site $s$.

\subsection{Decision Variables and Linear Constraints}
\indent To describe the errors in our noise channel, we introduce two binary error variables $y^e_{\ell},y^{\bar e}_{\ell}\in\{0,1\}$ for each oriented qutrit $\ell$, corresponding to the two possible orientations of $e\bar e$ pair creation. The variable $y^e_{\ell}=1$ denotes an error with orientation label $e$ on qutrit $\ell$, while $y^{\bar e}_{\ell}=1$ denotes an error with orientation label $\bar e$. Since the two orientations are mutually exclusive on each qutrit, we impose
\begin{equation}
	y^e_{\ell}+y^{\bar e}_{\ell}\leq 1.
	\label{z3_edge_constraint}
\end{equation}
As with Eq.~\ref{tc_single_pauli} for the $\mathbb{Z}_2$ TO under single-qubit Pauli noise, the constraint in Eq.~\ref{z3_edge_constraint} is imposed directly by the noise model but is automatically satisfied upon minimization of the objective function introduced below, and is therefore redundant.\\
\indent The fusion constraints enforce that, at each site $s$, the total anyon charge generated by errors on the support $\partial_+ s\cup\partial_- s$ fuses to the measured syndrome according to Eq.~\ref{fusion_Z3}. To express this condition as linear constraints, we introduce variables $n_s\in\{0,1,2\}$ that account for multiples of three, yielding
\begin{equation}
\sum_{\ell\in\partial_+ s}\left(y^e_{\ell}+2y^{\bar e}_{\ell}\right)+ \sum_{\ell\in\partial_- s}\left(2y^e_{\ell}+y^{\bar e}_{\ell}\right) =\sigma^{e}_s+2\sigma^{\bar e}_s +3n_s,
\label{z3_constraint}
\end{equation}
for each site $s$. The bound $n_s\in\{0,1,2\}$ is sufficient because, on the square lattice, the support of each site contains four qutrits, and the active error variable on each qutrit contributes at most the equivalent of two units of $e$-anyon charge to the site, so the left-hand side of Eq.~\ref{z3_constraint} is at most eight.
\subsection{Weights, Objective Function, and Threshold}
\indent As before, we choose the objective function as $f(\bm{x})=\ln P(E)$, with the trivial factor $P(\bm{\sigma}|E)$ enforced by the linear constraints. Under our noise channel, the creation of an $e\bar e$ pair in either orientation occurs with equal probability $\frac{p}{2}$ on each qutrit residing on link $\ell$. Therefore, all error variables have the same weight $w=\ln\frac{p/2}{1-p}<0$ and the objective function, equivalent up to an overall scaling by a nonzero constant, is
\begin{equation}
	f(\bm{x})=
	\sum_{\ell}
	\left(
	y^e_{\ell}
	+y^{\bar e}_{\ell}
	\right),
	\label{z3_objective}
\end{equation}
Minimum-weight decoding is obtained by minimizing $f(\bm{x})$.\\
\indent Numerically, this ILP decoder achieves a threshold of $15.346(5)\%$, close to the optimal threshold of $15.8(2)\%$ under the same noise channel~\cite{Z3_optimal, Z3-Simon-2026}. Furthermore, this ILP threshold is substantially higher than previously reported thresholds for renormalization group decoders, none of which exceed $13\%$~\cite{Z3_Poulin_2013, Z3_Anwar_2014, Z3_Hutter_2015}, as shown in Fig.~\ref{Fig:Z2Z3}.
\section{ILP for arbitrary anyon model}
\label{Sec:general_formulation}
\subsection{Input}
\indent In this section, we present the full ILP formulation for decoding arbitrary TOs, both Abelian and non-Abelian. For error correction with perfect anyon syndromes, the input to the ILP decoder consists of three components: the two-dimensional lattice whose sites support anyons, the measured anyon syndromes at those sites, and the fusion data of the TO.\\
\indent Anyons are assumed to occupy the sites of the lattice $S$. At each site $s\in S$, the anyon syndrome is determined by measuring a maximal set of locally compatible anyon species. As an example, Appendix~\ref{App:quantumdouble} details such a syndrome-measurement protocol for the quantum double model using commuting projectors \cite{KITAEV20032}. For now, we assume perfect anyon measurements, with measurement errors considered later in Sec.~\ref{measure_error}. While the lattice structure is fixed for a given realization of the TO, the syndrome information varies between decoder instances.\\
\indent The fusion rules of the TO are provided as input to the decoder and organized by fusion outcome. Let $\mathcal F_a$ denote the set of allowed fusion channels that produce an $a$-anyon at a lattice site. For each fusion channel $f\in\mathcal F_a$, let $N_{f,b}\in\mathbb Z_{\geq 0}$ denote the multiplicity of incident $b$-anyon strings associated with that channel. Since the connectivity at each site is typically a small finite integer, enumerating all allowed $f\in\mathcal F_a$ is feasible for relevant TOs. Examples of $f\in\mathcal F_a$ and the corresponding $N_{f,b}$ are illustrated in Fig.~\ref{Example}. The fusion rules are properties of the TO and are therefore also fixed for a given realization.\\
\indent In terms of the noise model, we consider physical errors that create incoherent pairs or multiplets of anyons on neighboring lattice sites, such that the anyon created at each site has a definite species label rather than being in a superposition of different species. When non-Abelian anyons created by different error events meet at a site, however, they may subsequently undergo nondeterministic fusion, producing a superposition of fusion outcomes prior to syndrome measurement. The noise model may also include correlations between incoherent creation events, either among different anyon species, as induced by Pauli $\hat{Y}$ errors in the $\mathbb{Z}_2$ TO, or across different lattice sites.
\subsection{Decision Variables and Linear Constraints}
\indent To capture the physical errors in the noise model, we introduce binary error variables $\epsilon_i\in\{0,1\}$ in one-to-one correspondence with local incoherent anyon-creation events. The index $i$ specifies both the location of the error and its anyon content. For example, in the $\mathbb{Z}_2$ TO, $i$ corresponds to the qubit label $\ell$ and the Pauli label $P$ of the error variable $y^P_{\ell}$. The value $\epsilon_i=1$ indicates that the corresponding error occurs, while $\epsilon_i=0$ indicates that it does not. These variables may be subject to constraints imposed directly by the noise model, such as those in Eq.~\ref{tc_single_pauli} or Eq.~\ref{z3_edge_constraint}. Because the error variables and their associated constraints are determined by the noise model, they remain fixed across decoder instances.\\
\indent Next, we introduce additional decision variables and constraints to ensure that the anyon strings specified by $\epsilon_i$ are consistent with both the fusion rules of the TO and the anyon syndromes measured at each lattice site. Since the syndromes vary between decoder instances, the corresponding fusion variables and constraints vary accordingly.\\
\indent At each site, the measurement yields a definite anyon label, which we denote by $a$; this may be the vacuum if all anyon syndromes are trivial. For each measured anyon species $a$ at site $s$, we introduce fusion variables $g_{f\to a,s}\in \{0,1\}$ for each fusion channel $f\in\mathcal F_a$, where $g_{f\to a,s}=1$ indicates that $a$ is produced through fusion channel $f$. Since the anyon strings are assumed to be incoherent, at most one fusion channel $f$ can be consistent with the given errors and measured syndrome. Accordingly, the variables $g_{f\to a,s}$ satisfy the linear constraint
\begin{equation}
    \sum_{f\in\mathcal F_a} g_{f\to a,s} = 1,
    \label{unique_fusion}
\end{equation}
which is imposed at each site $s$.\\
\indent Lastly, the multiplicities of anyons created by physical errors at site $s$ must be consistent with the selected fusion channel. If the measured syndrome at $s$ is labeled by $a$, then for each anyon species $b$, the number of incident $b$-anyons specified by the error configuration must equal the multiplicity of $b$ in the selected fusion channel $f\in\mathcal F_a$. This gives the linear constraint, for fixed $s$, $a$, and $b$,
\begin{equation}
    \sum_{\epsilon_i \mapsto b \text{ on } s} \epsilon_i = \sum_{f\in\mathcal F_a} N_{f,b} g_{f\to a,s},
    \label{fusion_constraint}
\end{equation}
where $\epsilon_i \mapsto b \text{ on } s$ indicates that the error $\epsilon_i$ incoherently creates a $b$-anyon at site $s$. This constraint is imposed for every anyon species $b$ at each lattice site $s$. For the Abelian $\mathbb{Z}_2$ and $\mathbb{Z}_3$ TOs under the noise channels considered above, Eqs.~\ref{unique_fusion} and \ref{fusion_constraint} reduce to Eqs.~\ref{tc_constraint} and \ref{z3_constraint}, respectively. In the general case involving non-Abelian fusion, however, Eqs.~\ref{unique_fusion} and \ref{fusion_constraint} explicitly keep track of the full fusion trees at each site.\\
\indent Collecting the decision variables, the vector $\bm{x}$ in Eq.~\ref{objective} is defined as 
\begin{equation}
	\bm{x}=\left(\bm{\epsilon}_{i}, \; \bm{g}_{f\to a,s}\right),
    \label{decision}
\end{equation}
whose dimension scales approximately linearly with the code distance multiplied by the number of anyon fusion channels in the TO. The number of linear constraints in Eqs.~\ref{unique_fusion}, and \ref{fusion_constraint} scales similarly. Since all decision variables in Eq.~\ref{decision} are binary, the formulation defines a BLP decoder, which is a special case of ILP. For specific TOs, the decision variables and constraints can often be reformulated in equivalent but simplified forms, thereby leading to faster runtime in practice. An example of such a reformulation is provided for the $D_4$ TO in Appendix~\ref{App:D4ILP} and is used to benchmark the performance of the ILP decoder in Sec.~\ref{Sec:D4}.
\subsection{Weights and Objective Function}
\indent As before, the objective function is chosen as $f(\bm{x})=\ln[P(\bm{\sigma}|E)P(E)]$, so that products of local probabilities are converted into additive contributions to the objective. The error variables $\epsilon_i$ enter through the prior probability $P(E)$. If each error $i$ occurs independently with probability $p_i$, activating $\epsilon_i$ contributes a factor $\frac{p_i}{1-p_i}$ to $P(E)$; accordingly, the associated weight is $w_i=\ln\frac{p_i}{1-p_i}$. When the noise model imposes additional constraints on the physical errors, such as mutual exclusivity, these weights should be adjusted accordingly, or effective weights may instead be used, as we do later for the $D_4$ TO. The fusion variables, by contrast, contribute to $P(\bm{\sigma}|E)$. At a site $s$ with measured anyon $a$, let $p_{f\to a}$ denote the probability that fusing the incident anyons through channel $f\in\mathcal F_a$ yields outcome $a$. The corresponding weight associated with $g_{f\to a,s}$ is then $w_{f\to a,s} = \ln p_{f\to a}$. Admittedly, for non-Abelian TOs, $P(\bm{\sigma}|E)$ is not fully captured by the product of the local factors $p_{f\to a}$. In particular, it also includes global consistency conditions requiring anyons created within isolated, homologically trivial components to fuse to the vacuum, which are difficult to encode using purely local probabilities. In practice, however, such conditions are often either automatically satisfied or can be relaxed, since the internal degrees of freedom of non-Abelian anyons that determine fusion outcomes are not accessible through syndrome measurements. Consequently, in most cases, the product of the local factors $p_{f\to a}$ over all sites still provides a good approximation to $P(\bm{\sigma}|E)$, and neglecting these global conditions has only a minimal effect on decoder performance.\\
\indent Therefore, the objective function can be written as 
\begin{equation}
    f(\bm{x})=\bm{w}^{\mathsf T} \bm{x}=\sum_{i}w_i \epsilon_i+\sum_s \sum_{f\in\mathcal F_a} w_{f\to a,s} g_{f\to a,s},
\end{equation}
so that the minimum-weight decoding, which is to find $\max_E P(\bm{\sigma}| E)P(E)$, is achieved by maximizing $f(\bm{x})$.

\section{Results for $D_4$ TO with Perfect Measurements}
\label{Sec:D4}
\indent To benchmark the performance of the ILP decoder, we numerically determine the error-correction threshold for the non-Abelian $D_4$ TO realized on a three-colorable kagome lattice \cite{Yoshida}. In this realization, the TO can be viewed as an Abelian $\mathbb Z_2^3$ TO that is ``twisted'' to become non-Abelian \cite{Yoshida, Iqbal2024, D4preparation}. The $D_4$ TO contains 22 anyon species generated by three Abelian anyons $e_c$ and three non-Abelian anyons $m_c$, with color label $c\in\{R,G,B\}$. The relevant fusion rules are
\begin{equation}
    m_R\times m_R = (1+e_G)(1+e_B), \quad e_R\times e_R = 1,
\end{equation}
together with permutations of the color labels. The $D_4$ TO is acyclic, since the non-Abelian $m$-anyon does not appear among its own fusion outcomes. In the kagome-lattice realization, the anyons are generated by physical Pauli errors. Specifically, Pauli $\hat X$ errors pair-create non-Abelian $m$-anyons of each color on the honeycomb lattice, whereas Pauli $\hat Z$ errors pair-create Abelian $e$-anyons of each color on the dual triangular lattice. We consider a single-qubit Pauli noise model in which all three colors of $m$- and $e$-anyons are generated independently, such that $p_y=p_xp_z$ on each physical qubit. This noise model generalizes that of Ref.~\onlinecite{jing2025intrinsicheralding}, in which only a single color of $m$-anyon is subject to noise, and is identical to the physical noise model considered in Ref.~\onlinecite{delafuente2026}. We assume perfect anyon measurements throughout.
\begin{figure}[t]
	\includegraphics[width=\columnwidth]{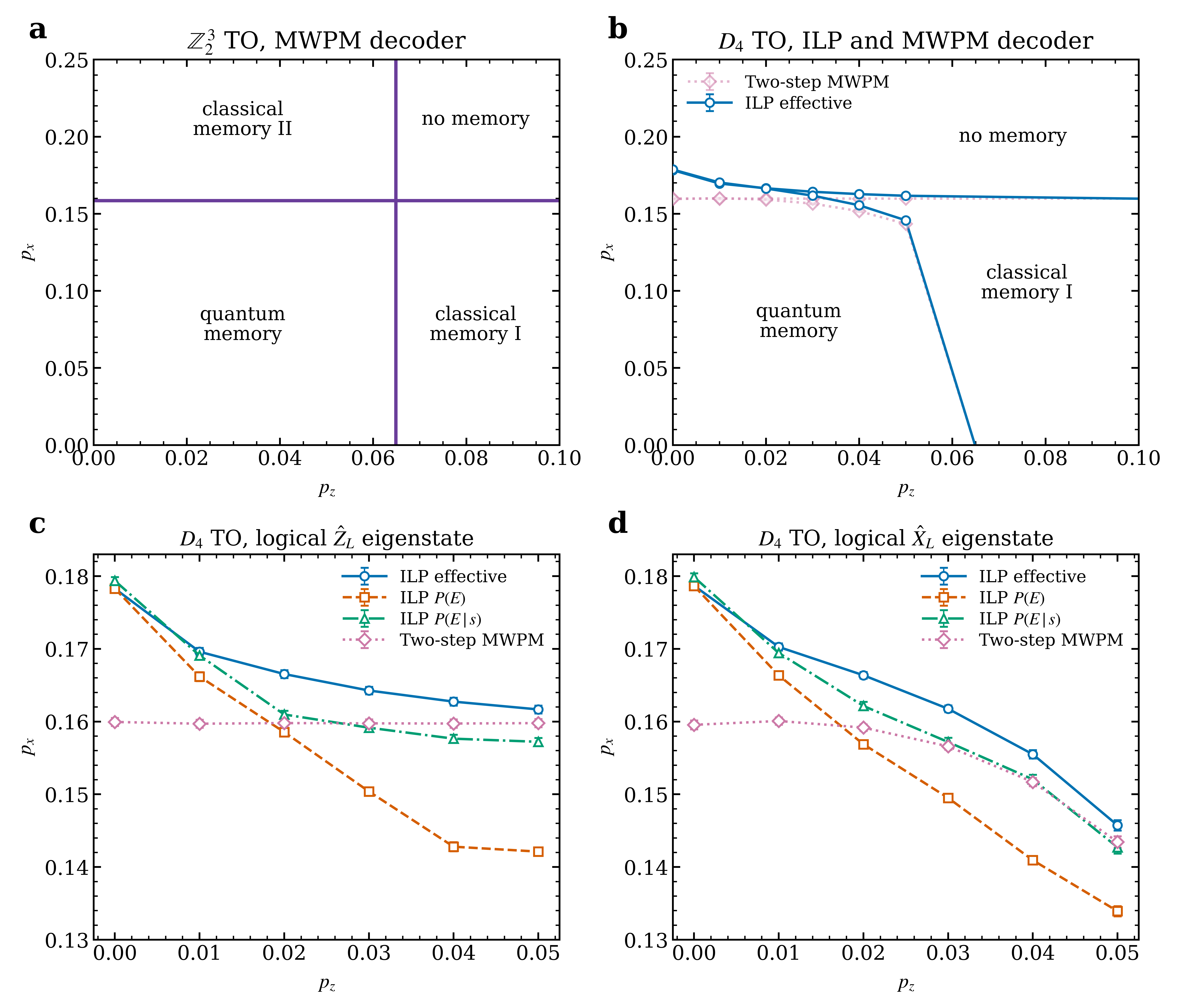}
	\caption{\textbf{Error-correction phase diagrams of the Abelian $\mathbb{Z}_2^3$ and non-Abelian $D_4$ TOs defined on the three-colorable kagome lattice.} (a) Error-correction phase diagram of the Abelian $\mathbb{Z}_2^3$ TO using the MWPM decoder. In addition to the quantum-memory and no-memory phases, there are two classical-memory phases, in which only the logical $\hat Z_L$ operator is protected (I) or only the logical $\hat X_L$ operator is protected (II). (b) Error-correction phase diagram of the non-Abelian $D_4$ TO obtained using the ILP decoder with an effective weight ratio and the two-step MWPM decoder. The classical-memory-II phase is absent because proliferation of the non-Abelian $m$-anyons necessarily entails proliferation of the Abelian $e$-anyons. (c) Error-correction phase boundary for the logical $\ket{0_L/1_L}$ states of the non-Abelian $D_4$ TO. (d) Error-correction phase boundary for the logical $\ket{+_L/-_L}$ states of the non-Abelian $D_4$ TO. In both (c) and (d), the ILP decoder that maximizes $P(E|\bm{\sigma})\propto P(\bm{\sigma}|E)P(E)$, shown by the green lines, outperforms the two-step MWPM decoder, shown by the pink lines, at small $p_z$ and achieves comparable thresholds at larger $p_z$. In contrast, the ILP decoder that considers only $P(E)$, shown by the orange lines, overestimates the heralding of $m$-anyons by $e$-anyons, improving the thresholds at small $p_z$ at the cost of substantially lower thresholds at large $p_z$. Finally, the ILP decoder with an effective weight ratio between $m$- and $e$-anyon strings achieves higher thresholds than all other decoders shown, establishing ILP as a better decoder for non-Abelian TOs with perfect syndrome measurements than currently available alternatives.}
	\label{D4ILP}
\end{figure}

\subsection{$D_4$ Phase Diagram and MWPM Decoding}
\indent Previous works have exploited the self-antiparticle and acyclicity properties of non-Abelian TOs to perform error correction using a two-step MWPM decoder \cite{BurtonIsing, davydova2026, jing2025intrinsicheralding, delafuente2026}. For the $D_4$ TO, the first step uses MWPM to correct the non-Abelian $m$-anyons, after which the Abelian $e$-anyons are measured and corrected in the second step. We first use this decoder to characterize the $D_4$ error-correction phase diagram and compare it with its Abelian $\mathbb{Z}_2^3$ counterpart. In the next subsection, we then benchmark the performance of the ILP decoder against MWPM.\\
\indent In Fig.~\ref{D4ILP}~(b), the pink dotted line shows the MWPM error-correction threshold for the single-qubit Pauli channel that independently generates all anyon species of the $D_4$ TO. For comparison, Fig.~\ref{D4ILP}~(a) shows the corresponding threshold for its Abelian counterpart, the $\mathbb{Z}_2^3$ TO, defined on the same lattice and evaluated using the same noise model and decoder \cite{MELCHERT20111828}. For the Abelian $\mathbb{Z}_2^3$ TO, the correction of $m$- and $e$-anyons decouples. Consequently, there are two error correction phases corresponding to a classical memory: one in which the logical $\hat Z_L$ operator is protected but the logical $\hat X_L$ operator is not (classical memory I), and another in which the logical $\hat X_L$ operator is protected but the logical $\hat Z_L$ operator is not (classical memory II). In contrast, as shown in Fig.~\ref{D4ILP}~(b), for the non-Abelian $D_4$ TO, the logical $\hat X_L$ operator is necessarily corrupted once the logical $\hat Z_L$ operator is no longer protected. Hence, only the classical memory I phase exists. This behavior follows from the fact that non-Abelian $m$-anyons can absorb $e$-anyons. The proliferation of $m$-anyons, which destroys the logical $\hat Z_L$ operator, therefore also induces the proliferation of $e$-anyons, which destroys the logical $\hat X_L$ operator. Similar behavior was observed for the steady state of the $D_4$ TO under a passive error-correction scheme \cite{chirame2024stabilizingnonabeliantopologicalorder}. The $D_4$ phase diagram with a single classical-memory phase is also observed for all three ILP decoder variants discussed below, indicating that this phase structure is a general feature of the $D_4$ TO rather than a consequence of the choice of decoder. Among these variants, ILP with an effective weight ratio achieves the highest error-correction threshold and is shown as the blue solid line in Fig.~\ref{D4ILP}~(b).

\subsection{Performance of ILP Decoders}
\indent Having established the MWPM benchmark and the structure of the phase diagram, we now study the application of the ILP decoder to the $D_4$ TO. Given perfect anyon syndromes, we consider three ILP decoder variants that differ only in their weight assignments. The first ignores $P(\bm{\sigma}|E)$ and maximizes $P(E)$; the second maximizes $P(\bm{\sigma}|E)P(E)$; and the third uses effective weights. Error correction is performed by applying both Pauli $\hat{X}$ and $\hat{Z}$ operators along the correction strings identified by the ILP decoder. Due to the nondeterministic fusion of non-Abelian anyons, a single application of the ILP decoder does not necessarily eliminate all anyons. However, the acyclicity of the $m$-anyons guarantees that all $m$-anyons are removed. After a subsequent round of $e$-anyon measurements, any remaining $e$-anyons can only lie along the applied Pauli $\hat{X}$ correction strings and should therefore be pairwise annihilated along the same strings. The simulation of the $D_4$ TO under Pauli noise, the generation of measurement outcomes, and the identification of logical errors after error correction are performed using the generalized stabilizer-tableau update method developed in Ref.~\onlinecite{chirame2024stabilizingnonabeliantopologicalorder}. Details of the MWPM decoding protocol and the three ILP decoding protocols are provided in Appendix~\ref{App:D4decoder}, while the numerical simulation details can be found in the public repository of Ref.~\onlinecite{ILP-Github}.\\
\indent In the first variant, the ILP decoder ignores the probabilistic collapse of the intermediate anyon superpositions created along non-Abelian anyon paths by setting $w_{f\to a,s}=0$ at all sites, and instead identifies the correction string that maximizes $P(E)$. As shown in Fig.~\ref{D4ILP}~(c) and (d), this method improves the thresholds for both logical $\ket{0_L/1_L}$ and logical $\ket{+_L/-_L}$ states (orange lines) beyond those of the two-step MWPM decoder (pink lines) at small values of $p_z$. However, the performance of the ILP decoder maximizing only $P(E)$ degrades sharply below that of the MWPM decoder as $p_z$ increases. This is because, at small $p_z$, Abelian $e$-anyons are primarily created along the error strings of non-Abelian $m$-anyons, and therefore provide useful intrinsic heralding for correcting $m$-anyons, even without accounting for $P(\bm{\sigma}|E)$ \cite{jing2025intrinsicheralding}. As $p_z$ increases, however, the heralding becomes unreliable and can incorrectly guide the correction of $m$-anyons, since $e$-anyons are predominantly created and moved by Pauli $\hat{Z}$ errors rather than appearing at sites where $m$-anyons fuse.\\
\indent Instead of considering only $P(E)$, the nondeterministic fusion encoded by $w_{f\to a,s}$ should also be included to construct the non-Abelian analog of minimum-weight decoding, which maximizes $P(E|\bm{\sigma})\propto P(\bm{\sigma}|E)P(E)$. Accounting for $P(\bm{\sigma}|E)$ introduces an additional penalty for $m$-anyon strings. Intuitively, this can be understood as an entropic factor, since longer $m$-anyon strings require repeated fusion into the channels specified by the intermediate syndrome measurements, each occurring with probability less than one at the corresponding site along the string. The resulting preference for shorter $m$-anyon strings effectively weakens the heralding provided by the $e$-anyons. As shown by the green lines in Fig.~\ref{D4ILP}~(c) and (d), the threshold improvement over the two-step MWPM decoder persists at small $p_z$, and the thresholds are slightly better than those of the ILP decoder that maximizes only $P(E)$ even if the heralding is weakened. At larger $p_z$, however, this decoder performs much better than the decoder based only on $P(E)$ and achieves thresholds close to those of the two-step MWPM decoder. As the heralding becomes increasingly unreliable with increasing $p_z$, it becomes difficult for any decoder to significantly outperform two-step MWPM, which does not use the $e$-anyon information. Indeed, in the limit of large $p_z$, the ILP decoder becomes equivalent to MWPM for the $m$-anyons.
\begin{figure*}[t]
	\includegraphics[width=1\textwidth]{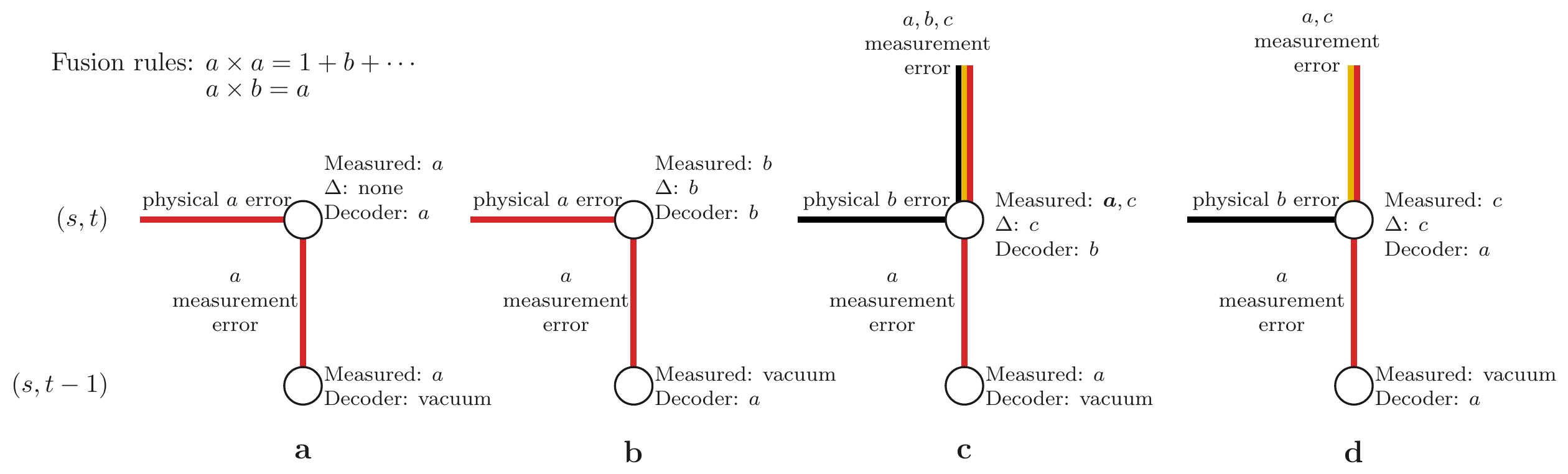}
	\caption{\textbf{Examples of fusion channels for spacetime ILP decoding.} (a) At the previous time step, a measurement error misidentified the vacuum as an $a$ anyon. At the current time step, assuming that no anyon strings of other species are present and no measurement error occurs, the absence of a defect $\Delta$ implies that an $a$ anyon is created by a physical error at $(s,t)$. This configuration is analogous to standard spacetime matching of defects in the Abelian $\mathbb{Z}_2$ TO \cite{Dennis}. (b) An $a$ anyon was not identified at the previous time step due to a measurement error and then fused with another $a$ anyon created by a physical error at time $t$, leading to a probabilistic syndrome measurement at $(s,t)$ that yields syndrome $b$ and contributes a nontrivial ILP weight. (c) At $(s,t-1)$, a measurement error reports a false-positive $a$ anyon, implying that the true syndrome is the vacuum. At $(s,t)$, multiple anyons are reported due to measurement errors. The ILP decoder may identify both reported anyons as false positives while assigning the true syndrome at $(s,t)$ to be a $b$ anyon created by a physical error. Since this $b$ anyon is present but unreported, an outgoing temporal $b$ string corresponding to the measurement error must also be present. (d) Although the incoming temporal string, spatial anyon strings, and defect at $(s,t)$ are the same as in (c), the absence of a reported $a$ syndrome at $(s,t)$ implies that an $a$ anyon is physically present but not identified due to a measurement error. By the fusion rule $a\times b=a$, the $b$ anyon created by a physical error can fuse with the $a$ anyon and go unreported without any additional measurement error, so no outgoing temporal $b$ string is required. Comparison with (c) illustrates why both the defect $\Delta$ and the measured syndrome are required to specify the fusion channel.}
	\label{SpacetimeExample}
\end{figure*}

\indent Weaker heralding improves error correction at larger $p_z$, whereas stronger heralding works well at smaller $p_z$. To get the best of both worlds, one can tune the relative weights of $m$- and $e$-anyon strings according to the physical error parameters, which can typically be learned on practical quantum devices. For simplicity, we set $w_{f\to a,s}=0$ and introduce a single effective parameter $r_{\mathrm{eff}}=w_m/w_e$. Since the ILP solution is unchanged by an overall rescaling of all weights, it is sufficient to consider only this ratio. For each noise parameter, we determine one optimal value of $r_{\mathrm{eff}}$ that yields the highest threshold. Because the ILP decoder makes discrete decisions, the optimal $r_{\mathrm{eff}}$ typically lies within a finite interval rather than at a single fine-tuned value. This makes the optimization straightforward and does not require advanced machine-learning techniques. The optimal values of $r_{\mathrm{eff}}$ obtained from our numerical simulations are provided in the repository of Ref.~\onlinecite{ILP-Github}. As shown by the blue lines in Fig.~\ref{D4ILP}~(c) and (d), the ILP decoder with this effective parameter outperforms both the two-step MWPM decoder and the other two ILP variants across the full range of noise parameters considered. Taken together, these results demonstrate that the ILP decoder provides a clear advantage for decoding non-Abelian TOs with perfect anyon syndromes.
\section{Measurement errors}
\label{measure_error}
\subsection{ILP Input}
\indent In the presence of anyon-syndrome measurement errors, multiple rounds of measurements are required to reliably infer the measurement errors before applying error correction \cite{Dennis}. Accordingly, the input lattice for the ILP decoder becomes three-dimensional, with spacetime sites $(s,t)$, where $s$ labels the spatial location and the additional time coordinate $t$ labels the syndrome-measurement round. The goal of the decoder is to infer the error configuration within a finite time interval during continuous error correction from the anyon syndromes measured over the corresponding rounds. Since errors may occur both before and after this interval, we impose open boundary conditions in the time direction.\\
\indent At each spacetime site $(s,t)$, the anyon syndrome $\bm a$ is determined by measuring a maximal set of locally compatible anyon species, such as the commuting projectors in the quantum double model described in Appendix~\ref{App:quantumdouble}. Due to measurement errors, multiple anyon species $\bm a$ may be reported at a given site, as illustrated in Fig.~\ref{SpacetimeExample}(c). Since at most one anyon species can physically occupy the site, at most one reported species can be correct, while the others are false-positive measurement outcomes.\\
\indent An error creating an anyon at site $(s,t)$ is detected through a change in the syndrome relative to the previous time step $(s,t-1)$, which we refer to as a defect $\Delta$. The full set of defects ${\Delta}$ within the time interval is therefore provided as input to the decoder. For Abelian TOs, defects alone are sufficient to infer error configurations \cite{Dennis}. An analogous situation is illustrated in Fig.~\ref{SpacetimeExample}(a), where the particular non-Abelian error configuration can likewise be inferred from the defects alone. The same is true for Fig.~\ref{SpacetimeExample}(b), which involves the non-Abelian fusion rule $a \times a = 1 + b + \cdots$. In general, however, for non-Abelian TOs, the measured syndrome $\bm a$ at each site must also be provided in addition to the defects to determine the allowed spacetime anyon-string configurations, as illustrated in Fig.~\ref{SpacetimeExample}(c,d), which show the same defects but different syndrome outcomes. For example, if the fusion rule includes $a\times b=a$, then a physical error can create a $b$ anyon at a site already containing an $a$ anyon without producing a defect.\\
\indent The fusion rules of the TO are provided as input to the decoder through the set of allowed fusion channels, denoted by $\mathcal F_{\bm a,\Delta}$, that are consistent with the measured syndrome $\bm a$ and defect $\Delta$ at spacetime site $(s,t)$. For each fusion channel $f\in\mathcal F_{\bm a,\Delta}$, let $N^{\mathrm{sp}}_{f,b}\in\mathbb Z{\geq 0}$ denote the multiplicity of incident spatial $b$-anyon strings created by physical errors, while $N^{\mathrm{in}}_{f,b},N^{\mathrm{out}}_{f,b}\in\{0,1\}$ denote the multiplicities of incoming and outgoing temporal $b$-anyon strings corresponding to measurement errors at sites $(s,t-1)$ and $(s,t)$, respectively. In the absence of measurement errors, the allowed fusion channels $f\in\mathcal F_{\bm a,\Delta}$ reduce to the set $\mathcal F_a$ introduced in Sec.~\ref{Sec:general_formulation}. Examples of these fusion channels and the corresponding $N^{\mathrm{sp}}_{f,b}$, $N^{\mathrm{in}}_{f,b}$, and $N^{\mathrm{out}}_{f,b}$ are illustrated in Fig.~\ref{SpacetimeExample}.\\
\indent In terms of the noise model, we consider ILP decoding against physical errors that create incoherent, possibly correlated, pairs or multiplets of anyons on neighboring sites, together with anyon-syndrome measurement errors, which are inherently incoherent. For simplicity, we assume that physical and measurement errors are uncorrelated.
\subsection{Decision Variables, Constraints, and Weights}
\indent As in Sec.~\ref{Sec:general_formulation}, we introduce binary error variables $\epsilon_i\in\{0,1\}$ for physical errors. These variables are in one-to-one correspondence with local incoherent anyon-creation events, with the index $i$ specifying both the spacetime location of the error and its anyon content. They may also be subject to constraints imposed directly by the noise model. In addition, we introduce binary variables $m_{a,s,t}\in\{0,1\}$ for $a$-anyon measurement errors at site $(s,t)$. For both types of error variables, a value of $1$ indicates that the corresponding error occurs.\\
\indent Then, for each spacetime site $(s,t)$ with measured syndrome $\bm a$ and defect $\Delta$, we introduce a binary fusion variable $g_{f\to(\bm a,\Delta),s,t}\in\{0,1\}$ for each allowed fusion channel $f\in\mathcal F_{\bm a,\Delta}$. The decoder is therefore a BLP, with the decision-variable vector $\bm{x}$ in Eq.~\ref{objective} given by
\begin{equation}
	\bm{x}=\left(\bm{\epsilon}_{i}, \; \bm{m}_{a,s,t}, \; \bm{g}_{f\to\left(\bm a,\Delta\right),s,t}\right).
    \label{3Ddecision}
\end{equation}
\indent The multiplicities of incident spatial and temporal $b$-anyon strings must be consistent with the selected fusion channel $f\in\mathcal F_{\bm a,\Delta}$. For each anyon species $b$ and spacetime site $(s,t)$ with measured syndrome $\bm a$ and defect $\Delta$, this requirement gives separate constraints for the spatial, outgoing temporal, and incoming temporal $b$ anyon strings:
\begin{align}
    \sum_{\epsilon_i \mapsto b \text{ on } (s,t)} \epsilon_i &= \sum_{f\in \mathcal F_{\bm a,\Delta}} N^{\mathrm{sp}}_{f,b}\, g_{f\to\left(\bm a,\Delta\right),s,t}, \nonumber\\
    m_{b,s,t} &= \sum_{f\in \mathcal F_{\bm a,\Delta}} N^{\mathrm{out}}_{f,b}\, g_{f\to\left(\bm a,\Delta\right),s,t},\\
    m_{b,s,t-1} &= \sum_{f\in \mathcal F_{\bm a,\Delta}} N^{\mathrm{in}}_{f,b}\, g_{f\to\left(\bm a,\Delta\right),s,t}, \nonumber
    \label{3Dfusion_constraint}
\end{align}
which generalize Eq.~\ref{fusion_constraint} of Sec.~\ref{Sec:general_formulation} to spacetime decoding with measurement errors.\\
\indent Lastly, the unique choice of $f\in\mathcal F_{\bm a,\Delta}$ at each site $(s,t)$ is enforced by
\begin{equation}
    \sum_{f\in \mathcal F_{\bm a,\Delta}} g_{f\to\left(\bm a,\Delta\right),s,t} = 1,
\end{equation}
analogous to Eq.~\ref{unique_fusion} in Sec.~\ref{Sec:general_formulation}.\\
\indent With the decision variables subject to the linear constraints above, the BLP decoder identifies the most likely error configuration by maximizing the linear objective function
\begin{align}
    f(\bm{x})=\bm{w}^{\mathsf T} \bm{x}=&\sum_{i}w_i \epsilon_i+\sum_{s,t}\sum_a w_{a,s,t} m_{a,s,t} \nonumber\\
    &+\sum_{s,t}\sum_{f\in\mathcal F_{\bm a,\Delta}} w_{f,s,t} g_{f,s,t}.
\end{align}
Here, $w_i=\ln\frac{p_i}{1-p_i}$ accounts for the probability $p_i$ of the physical error $\epsilon_i$ under the assumption that physical errors occur independently, and should be adjusted when additional constraints are imposed by the noise model. Similarly, $w_{a,s,t}=\ln\frac{q_{a,s,t}}{1-q_{a,s,t}}$ accounts for the probability $q_{a,s,t}$ of an $a$-anyon measurement error at site $(s,t)$. Given the selected fusion channel $f\in\mathcal F_{\bm a,\Delta}$, the decoder infers how the anyons created by physical errors along the spatial anyon strings fuse with the true anyon from the previous time step, as determined by $\bm a$, $\Delta$, and the incoming temporal strings. The resulting fusion outcome, inferred from $\bm a$ and the outgoing temporal strings, is obtained upon measurement with probability $p_{f\to(\bm a,\Delta)}$ according to the fusion rules of the TO and contributes the weight $w_{f,s,t}=\ln p_{f\to(\bm a,\Delta)}$ to the objective. As in Sec.~\ref{Sec:general_formulation}, although global consistency conditions are not accounted for, this weight assignment nevertheless provides a good approximation to $P(E|\bm{\sigma})\propto P(\bm{\sigma}|E)P(E)$ for minimum-weight decoding.
\subsection{Just-in-time ILP}
\indent The ILP decoder described in the previous subsections determines the minimum-weight error configuration within a finite spacetime region from the measured anyon syndromes. For Abelian TOs, continuous error correction can be performed by applying one round of ILP decoding over the entire code after $T=\mathcal O(d)$ rounds of measurements \cite{Dennis}. For non-Abelian TOs, however, the fusion rules allow anyons to fuse into other non-Abelian anyons. Consequently, accumulated non-Abelian charge can hide additional anyon content, making delayed decoding less reliable. It is therefore preferable to apply corrections as soon as measurement errors can be inferred with sufficient confidence, i.e., just in time \cite{bombinJIT, BrownJIT, ScrubyJIT, davydova2026, delafuente2026}. Although the just-in-time strategy is not necessary for Abelian TOs, it can also be applied to them. We therefore incorporate just-in-time techniques into the spacetime ILP decoder, enabling continuous error correction for arbitrary TOs in the presence of anyon-syndrome measurement errors. Our protocol largely follows the matching-based just-in-time algorithms proposed in Ref.~\onlinecite{davydova2026} for the $D_4$ TO, which rely on its self-antiparticle and acyclic fusion properties, while replacing the matching decoder with our ILP decoder.
\begin{algorithm}[H]
  \caption{Just-in-time ILP decoder}
  \label{algorithm}
  Initialize an empty set $\bm{\Delta}$ to collect all uncorrected defects. At each time step $t$:
   \begin{itemize}
   \item[1:] Measure and record the anyon syndromes at time $t$.
   \item[2:] Compute the defects $\Delta$ relative to the syndrome history at time $t-1$ and add all nontrivial defects to $\bm{\Delta}$.
   \item[3:] Perform ILP decoding using the syndrome history and the uncorrected defects $\bm{\Delta}$. In the ILP solution, defects in $\bm{\Delta}$ are connected by correction strings either to one another or to the current time boundary, thereby partitioning them into disconnected clusters. For each cluster, let $Q$ denote the linear size of the smallest spacetime cube containing the cluster.
   \begin{itemize}
       \item[3.1] If the cluster contains any defect or correction string younger than $Q$, i.e., occurring after time $t-Q$, leave the cluster uncorrected at the current time step.
       \item[3.2] Otherwise, apply correction circuits that move all defects in the cluster from their spatial locations to the spatial center of the cluster, and remove all nontrivial defects in the cluster from $\bm{\Delta}$.
   \end{itemize}
   \end{itemize}
\end{algorithm}
\indent Colloquially, the ILP decoder uses the syndrome history to identify clusters of uncorrected defects. Before correcting a cluster at the current time step, we require its age to exceed its spatial and temporal extent, so that measurement errors can be inferred with sufficient confidence. The defects within a cluster can then be corrected by applying exact anyon strings. Although such strings are generally of linear depth \cite{Beckman02,Shi19,Bravyi22,Liu22}, their depth can remain comparable to that of the syndrome-measurement circuit in the subthreshold regime, where the clusters are small, making such correction between successive measurement rounds reasonable. Alternatively, one may apply the inverse of the inferred error operators, which can be implemented in finite depth for local physical noise channels. In the proposed algorithm, the ordering of correction operations, corresponding to different braiding processes, is not explicitly tracked. This is because braiding affects the internal degrees of freedom of non-Abelian anyons, which are not accessible through the syndrome measurements provided to the decoder. After correction, the cluster is removed from the set of uncorrected defects, while any nontrivial fusion outcomes resulting from the correction are effectively pushed to the current time step $t$.\\
\indent To prove the existence of a threshold, one typically considers hierarchically defined error clusters, which become increasingly rare at higher levels of the hierarchy, and shows that error correction does not spread a cluster into well-separated larger clusters \cite{HarringtonThesis, Dauphinais2017, davydova2026, lyons2026}. Since this proof-of-principle algorithm does not increase the spatial extent of an identified cluster in the absence of larger nearby clusters, it is plausible that a threshold proof could be obtained by further showing that correction does not spread the cluster temporally into higher levels of the hierarchy. We leave the explicit construction and numerical study of a just-in-time ILP decoder for practical non-Abelian TOs, together with a threshold proof, to future work.
\section{Conclusion and Outlook} 
\indent In this work, we introduced an integer linear programming (ILP) framework for decoding arbitrary TOs, both Abelian and non-Abelian. By introducing auxiliary variables, the framework linearizes the error-correction problem and enables minimum-weight decoding using classical optimization methods. Through linear constraints, the decoder captures correlated errors and the fusion rules of arbitrary TOs, which are generally inaccessible to matching-based or clustering decoders. As demonstrated for the Abelian $\mathbb{Z}_2$ and $\mathbb{Z}_3$ TOs and the non-Abelian $D_4$ TO, the ILP decoder achieves substantially improved error-correction thresholds under incoherent anyon noise compared with most existing decoders. Looking ahead, the underlying optimization may be further accelerated using modern machine learning techniques. Moreover, because the ILP optimization can admit degenerate optima, namely distinct error configurations with the same value of $f(\bm{x})$, improved tie-breaking strategies may further enhance decoder performance. Possible approaches include methods inspired by ordered statistics decoding \cite{splitbelief, quantumBP1, quantumBP2, gu2025linearprogramming}.\\
\indent An exciting direction is to apply the ILP framework to a broader range of physically and computationally interesting anyon models and noise channels. Recent work has analyzed the phase diagram of the $D_4$ TO considered here when particular types of non-Abelian anyons proliferate simultaneously (namely $m_R$ and $m_B$), finding that their proliferation can parasitically condense a shared Abelian fusion outcome (namely $e_G$) \cite{Vadali2026}. Because such proliferation is a special case of our Pauli noise model, one could revisit this phase diagram and its anyon proliferation with the ILP decoder. We note that in Fig.~\ref{D4ILP} we already observed that proliferating all three non-Abelian anyons (i.e., $m_R$, $m_B$, and $m_G$) leads to a completely trivial phase, and it would be interesting to understand the nature of this transition and whether it involves parasitic condensation of the Abelian anyons. Further afield, applying the ILP framework to decoding Fibonacci TO would permit direct comparison with existing clustering decoders that incorporate information about the fusion and braiding of Fibonacci anyons \cite{BurtonFibonacci, Verstraete}. It would be equally interesting to study error correction in the presence of proliferating fermions in the toric code \cite{Wang2025IntrinsicMixed,ChenGrover2024Coherent}.\\
\indent A broader question concerns the input to the decoder. Recent information-theoretic studies characterize recovery thresholds of decohered non-Abelian TOs without restricting to a specific syndrome measurement protocol \cite{Pablo1,Pablo2,kimmcginley,Kim2026OptimalRecovery}. These results raise the possibility that local anyon-syndrome measurements may not always retain all information relevant for recovery, motivating collective or even adaptive measurements and their integration with optimization-based decoders. This issue becomes especially important for continuum and chiral topological phases, which generally lack a commuting set of local syndrome observables. For example, recent work on fractional quantum Hall states under density decoherence finds that information encoded in the fusion space of anyons in the Moore-Read state can remain recoverable even under strong decoherence \cite{WangAltman2025FQH}. Although formulating experimentally accessible measurement protocols and correction operations in such systems is considerably more open-ended, the broader idea underlying the ILP approach, combining available measurement outcomes, the physical noise model, and fusion rules in a global linear optimization formulation, may remain useful beyond lattice models with conventional anyon syndromes.\\
\indent Beyond minimum-weight decoding via deterministic optimization, the ILP formulation establishes a direct connection between error correction and classical statistical mechanics. The binary decision variables in the general formulation of Sec.~\ref{Sec:general_formulation}, which encode anyon-string configurations, may be viewed as classical Ising degrees of freedom. Because the constraints in the general formulation are equalities, the admissible anyon-string configurations form a constrained subspace of the full configuration space spanned by all decision variables. In principle, an optimal decoder could therefore be obtained by sampling this constrained subspace according to the weights $\bm{w}_{i}$: admissible configurations are connected ergodically by local updates, and their relative probabilities can be evaluated locally. Although the complexity of Monte Carlo sampling schemes can scale polynomially with code distance, such schemes are expected to be slow in practice. It therefore remains an open question whether they provide a realistic route to optimal decoding for arbitrary TOs with anyon-syndrome measurements. If so, the critical behavior of the corresponding optimal error correction phase transition would be an interesting direction for future work.\\
\indent The ILP framework also extends to decoding in the presence of measurement errors and therefore provides a promising route toward spacetime decoding. Although we present the framework and formulate a just-in-time ILP decoder, its numerical implementation and performance benchmarking remain for future work. Furthermore, a more careful definition of error clusters in Algorithm~\ref{algorithm} may provide a route toward proving the existence of an ILP decoding threshold in the presence of measurement errors.
\section*{Acknowledgements}
\indent D.J. would like to acknowledge Sanket Chirame for explaining the stabilizer tableau formalism for the $D_4$ TO, Tianfu Wang for discussions on linear programming, Anasuya Lyons for discussions on just-in-time decoding, Pablo Sala for discussions on non-Abelian error correction, and Ramanjit Sohal for discussions on non-Abelian topological orders. A.Z. was supported by the Selove Summer Research Prize, awarded by the Department of Physics at the University of Chicago. We acknowledge support from the ARO(W911NF-23-1-0077), ARO MURI (W911NF-21-1-0325), AFOSR MURI (FA9550-21-1-0209, FA9550-23-1-0338), ONR MURI (N000142612102), DARPA (HR0011-24-9-0361), NSF (ERC-1941583, OMA-2137642, OSI-2326767, CCF-2312755, OSI-2426975). This material is based upon work supported by the U.S. Department of Energy, Office of Science, National Quantum Information Science Research Centers and Advanced Scientific Computing Research (ASCR) program under contract number DE-AC02-06CH11357 as part of the InterQnet quantum networking project. This work was completed with resources provided by the University of Chicago’s Research Computing Center.

\bibliographystyle{apsrev4-2}
\bibliography{references}

\appendix
\onecolumngrid

\newpage
\section{ILP Runtime for the Abelian $\mathbb{Z}_2$ TO under Single-Qubit Pauli Noise}
\label{App:Z2}
\indent In this appendix, we characterize the runtime of the ILP decoder for the Abelian $\mathbb{Z}_2$ TO under single-qubit Pauli noise. Although the minimum-weight decoding problem in this setting is NP-hard~\cite{NPhard1, NPhard2, NPcomplete}, and polynomial-time scaling therefore cannot be guaranteed in the worst case, worst-case asymptotic complexity does not necessarily determine the runtime for physically relevant error configurations. We therefore study the distribution of solver runtimes and their finite-size behavior. We find three main features. First, for a fixed code distance $d$ and error rate $p$, the runtime distribution develops a pronounced upper tail, so the mean runtime is strongly influenced by the hardest error configurations. Second, below the error-correction threshold, where quantum error-correcting codes typically operate, both the mean and median runtimes are consistent with polynomial scaling in the number of physical qubits, $N=2d^2$ for the square-lattice realization used in the main text, over the numerically accessible code distances. Third, although the runtime increases substantially near the error-correction threshold, its dependence on $N$ remains inconsistent with a simple exponential dependence over the accessible range. Thus, these results indicate that the ILP decoder for the Abelian $\mathbb{Z}_2$ TO under single-qubit Pauli noise can operate efficiently over practically relevant error rates and system sizes.
\subsection{Runtime Statistics}
\begin{figure}[b]
    \centering
    \includegraphics[width=0.8\textwidth]{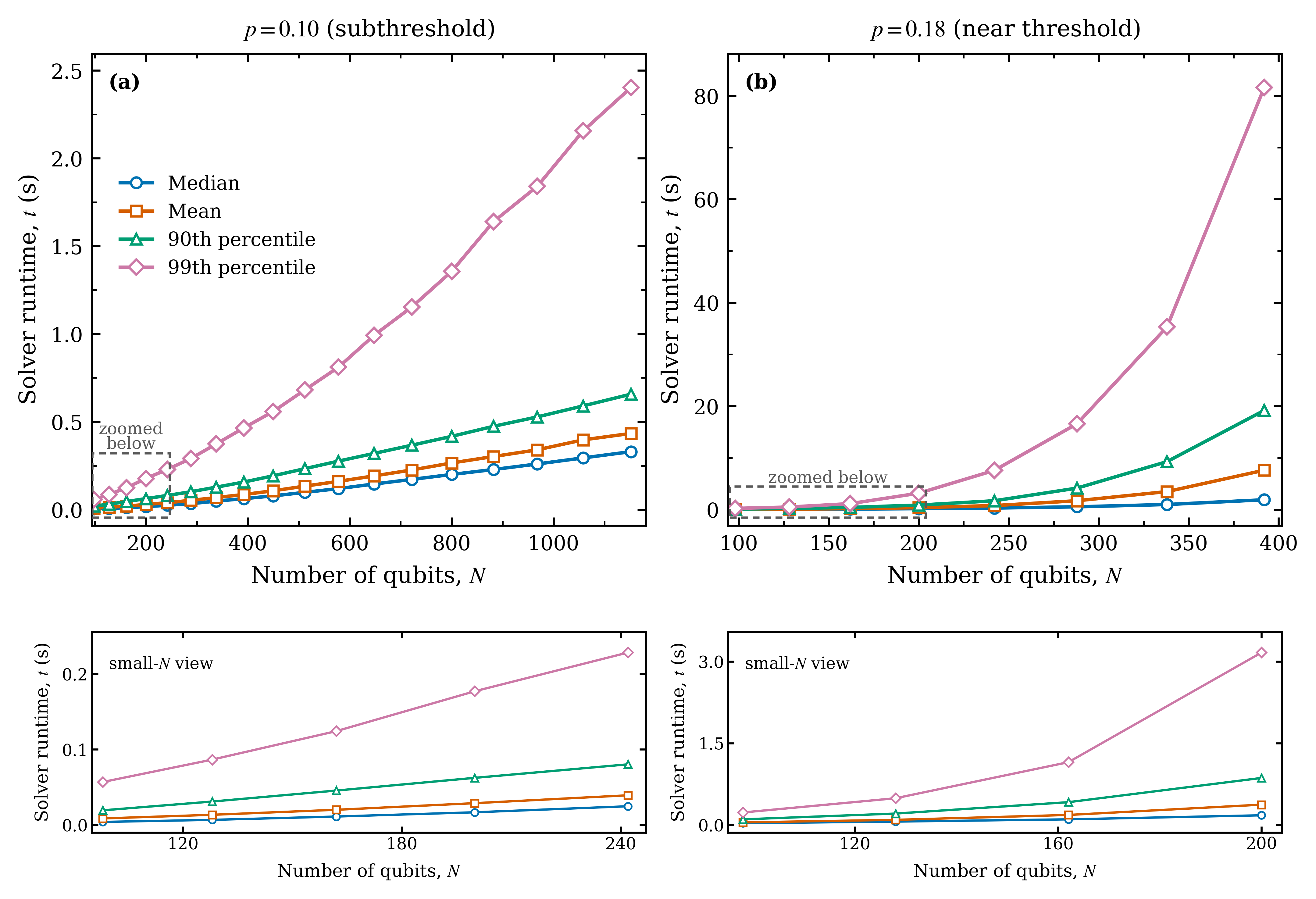}
    \caption{\textbf{Runtime distribution and upper-tail behavior.}
    Median, mean, 90th-percentile, and 99th-percentile optimization times versus the number of physical qubits $N=2d^2$ at (a) the subthreshold error rate $p=0.10$ and (b) $p=0.18$, near the error-correction threshold. The lower panels enlarge the small-$N$ region. The growing separation among the median, mean, and upper percentiles demonstrates substantial decoder-instance variability and a pronounced upper tail. The 99th percentile characterizes this tail and should not be interpreted as a worst-case runtime.}
    \label{fig:runtime_distribution}
\end{figure}
\indent We study the runtime of the ILP decoder at a subthreshold error rate $p=0.10$ for $d=7$ ($N=98$) to $d=24$ ($N=1152$), and near the error-correction threshold, at $p=0.18$, for $d=7$ ($N=98$) to $d=14$ ($N=392$). The reported runtime is the wall-clock time spent in the classical optimization used to determine the minimum-weight error configuration and excludes the construction of the ILP model. The data in Fig.~\ref{fig:runtime_distribution} show the median, mean, 90th percentile, and 99th percentile of the runtime obtained from $10^5$ independently sampled physical error configurations for each code distance at $p=0.10$ and $10^4$ configurations for each code distance at $p=0.18$. As shown in Fig.~\ref{fig:runtime_distribution}, the runtime varies widely among decoder instances, with the distribution broadening as $N$ increases and causing the mean and upper percentiles to become increasingly sensitive to rare, difficult instances, especially near the threshold.
\subsection{Subthreshold Runtime Scaling}
\begin{figure}[t]
    \centering
    \includegraphics[width=\textwidth]{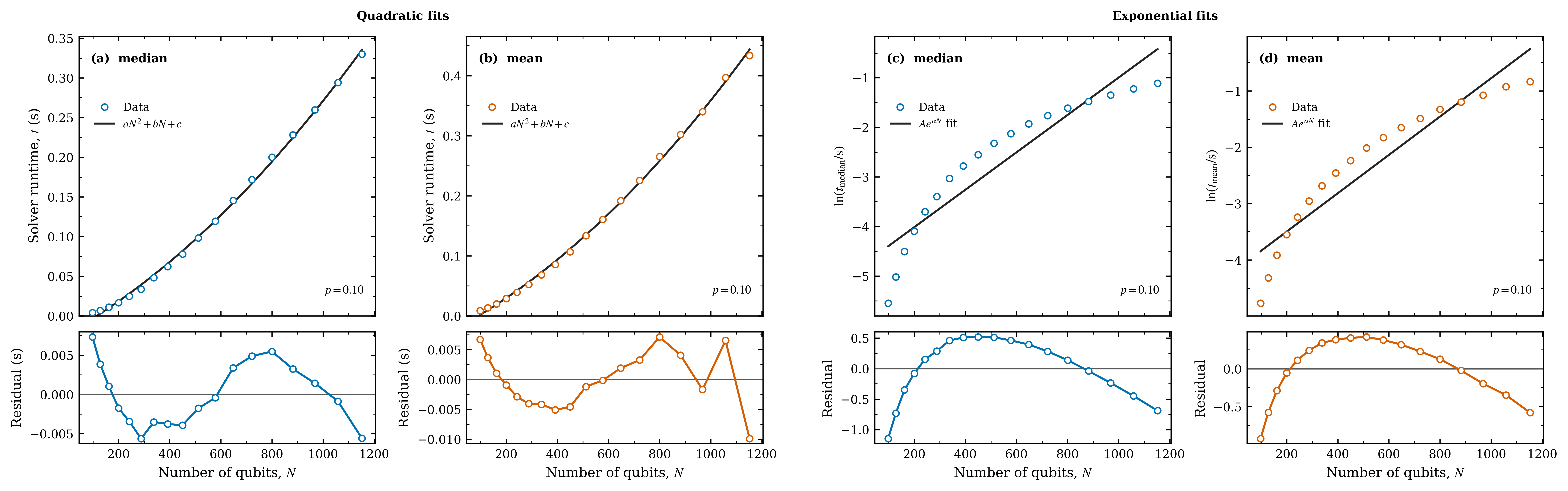}
    \caption{\textbf{Subthreshold runtime scaling.} Panels (a) and (b) show the median and mean solver runtime, respectively, versus $N=2d^2$ at $p=0.10$, together with quadratic fits $t(N)=aN^2+bN+c$. For the median, $a=1.1264\times10^{-7}$, $b=1.8064\times10^{-4}$, and $c=-2.2205\times10^{-2}$; for the mean, $a=1.5309\times10^{-7}$, $b=2.2772\times10^{-4}$, and $c=-2.1999\times10^{-2}$. Both statistics are well described by the quadratic model over the numerically accessible system sizes. Panels (c) and (d) show the same median and mean runtime data, respectively, on log-linear axes, together with fits to the simple exponential form $t(N)=Ae^{\alpha N}$. The clear systematic deviations from the fitted curves show that a simple exponential dependence on $N$ provides a poor description of the data. The lower panels show the residuals of the corresponding quadratic and exponential fits.}
    \label{fig:runtime_subthreshold}
\end{figure}
\begin{figure}[b]
    \centering
    \includegraphics[width=\textwidth]{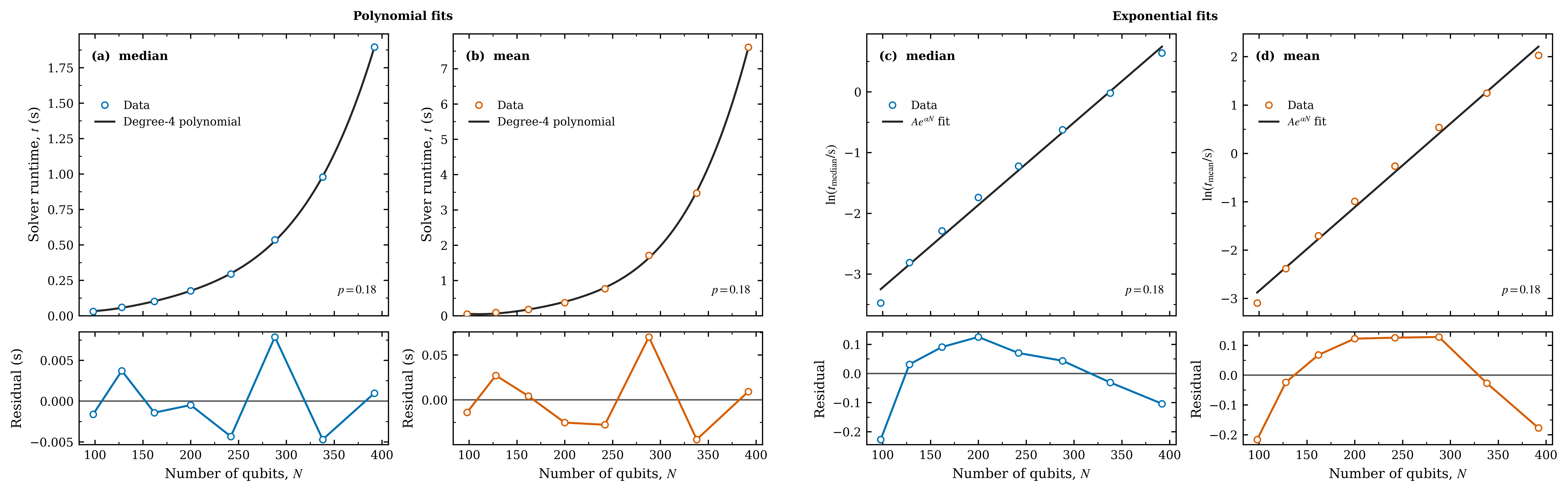}
    \caption{\textbf{Runtime scaling near threshold.} Panels (a) and (b) show the median and mean solver runtime, respectively, at $p=0.18$ versus the number of physical qubits $N=2d^2$, together with degree-four polynomial fits. Panels (c) and (d) show the same median and mean runtime data, respectively, on log-linear axes, together with fits to the simple exponential form $t(N)=Ae^{\alpha N}$. The lower panels show the residuals of the corresponding polynomial and exponential fits. For the exponential fits, the pronounced systematic progression from negative to positive and back to negative residuals, clearly visible in both panels, is inconsistent with random scatter about a straight line, indicating that a simple exponential dependence on $N$ does not describe the accessible range.}
    \label{fig:runtime_threshold}
\end{figure}
\indent To study the runtime scaling below the error-correction threshold, we fit the median and mean runtimes at $p=0.10$ to second-order polynomials in the number of physical qubits. As shown in panels (a) and (b) of Fig.~\ref{fig:runtime_subthreshold}, the fits closely follow the data across all accessible code distances, indicating that both the median and mean runtimes are consistent with polynomial scaling in the subthreshold regime. Although the fitted polynomial degree should not be interpreted as evidence that the asymptotic runtime is quadratic, these results show that the NP-hardness of the error-correction problem does not preclude the ILP decoder from remaining computationally efficient over physically relevant subthreshold error rates and code distances.\\
\indent For comparison, panels (c) and (d) of Fig.~\ref{fig:runtime_subthreshold} show the same median and mean runtime data, respectively, on log-linear axes together with fits to a simple exponential dependence, which would appear as a straight line on these axes. The pronounced systematic deviations from the fitted lines make clear that a simple exponential form does not describe the runtime statistics over the accessible code distances.
\subsection{Scaling Near the Error-Correction Threshold}
\indent Near the threshold, we test for simple exponential runtime scaling, $t(N)=Ae^{\alpha N}$, by fitting straight lines to the median and mean runtime data on log-linear axes. As shown in panels (c) and (d) of Fig.~\ref{fig:runtime_threshold}, both statistics exhibit downward curvature, with residuals that vary systematically from negative to positive and back to negative. Thus, although the near-threshold runtime is substantially larger than the subthreshold runtime, its finite-size scaling is inconsistent with a simple exponential dependence $t\propto e^{\alpha N}$ over the numerically accessible code distances.\\
\indent For comparison, we also fit the near-threshold data to a fourth-degree polynomial, which closely follows the accessible data. This fit should, however, be interpreted cautiously, since it contains five free parameters for only eight data points, leaving just three residual degrees of freedom.

\section{Anyon Syndrome Measurements in Quantum Double Models as Input to the ILP Decoder}
\label{App:quantumdouble}
\indent In this appendix, we discuss the measurement of a maximal set of locally compatible anyon species to obtain the anyon syndrome as input to the ILP decoder for arbitrary TOs. As a concrete example, we present the protocol in the context of the quantum double model $D(G)$ of a finite group $G$. The model is defined on an oriented two-dimensional lattice, with each link carrying a $|G|$-dimensional Hilbert space whose basis states are labeled by group elements. Anyon excitations are localized on sites $s=(u,p)$, where $u$ is a microscopic lattice site and $p$ is an adjacent plaquette.\\
\indent The anyon species of $D(G)$ are labeled by pairs $(\mathcal C,\rho)$. Here $\mathcal C$ is a conjugacy class of $G$, and $\rho$ is an irreducible representation of the centralizer $Z_{g_0}$ of a fixed representative $g_0\in\mathcal C$. The vacuum corresponds to the trivial conjugacy class together with the trivial representation. For a site $s=(u,p)$, a complete syndrome measurement is given by the mutually orthogonal projectors
\begin{equation}
	\Pi_s^{(\mathcal C,\rho)}
	=
	\sum_{g\in\mathcal C}
	B_p^g
	\left[
	\frac{d_\rho}{|Z_g|}
	\sum_{h\in Z_g}
	\chi_{\rho_g}(h^{-1})A_s^h
	\right],
	\label{qd_projector}
\end{equation}
where $B_p^g$ projects onto states with oriented plaquette product $g$, and $A_s^h$ denotes the local gauge transformation by $h$ at the microscopic lattice site $u$. For $g=x_g g_0 x_g^{-1}$, the representation $\rho_g$ of $Z_g=x_g Z_{g_0}x_g^{-1}$ is defined by $\rho_g(h)=\rho(x_g^{-1}h x_g)$, $d_\rho$ is the dimension of $\rho$, and $\chi_{\rho_g}$ is the character of $\rho_g$. These projectors satisfy
\begin{equation}
	\Pi_s^{(\mathcal C,\rho)}\Pi_s^{(\mathcal C',\rho')}
	=
	\delta_{\mathcal C,\mathcal C'}\delta_{\rho,\rho'}
	\Pi_s^{(\mathcal C,\rho)},
	\;\;\;
	\sum_{(\mathcal C,\rho)}\Pi_s^{(\mathcal C,\rho)}=\mathbb I_s,
\end{equation}
and commute with the corresponding projectors on other sites. Thus, perfect syndrome measurements assign a unique anyon label $(\mathcal C,\rho)$, including the vacuum label, to each site $s$, providing the syndrome input to the ILP decoder as discussed in Sec.~\ref{Sec:general_formulation}. When the commuting-projector measurements are imperfect, multiple anyon species may be reported at a given site $s$, although at most one of the reported species can be correct, as discussed in Sec.~\ref{measure_error}.

\section{ILP Formulation for Decoding the $D_4$ Topological Order under Single-Qubit Pauli Noise}
\label{App:D4ILP}
\indent In this appendix, we present an ILP formulation for decoding the non-Abelian $D_4$ TO under a single-qubit Pauli noise channel that generates all anyon species. This formulation is equivalent to the general formulation introduced in Sec.~\ref{Sec:general_formulation}, but uses fewer decision variables and constraints than a naive application of the general framework and is substantially faster in practice. We therefore use it in the numerical simulations that determine the error-correction threshold in Sec.~\ref{Sec:D4}.\\
\indent The ILP formulation presented here applies to any lattice realization of the non-Abelian $D_4$ TO. For concreteness, however, we consider the realization on a three-colorable kagome lattice defined by the quasi-stabilizer Hamiltonian
\begin{equation}
    H_{D_4}
    \;=\;
    -\sum_{s\in \{\textrm{\ding{65}}\}}A_s
    \;-\;
    \sum_{t\in \{\triangleright,\,\triangleleft\}} B_t, 
    \label{D4_Hamiltonian}
\end{equation}
where the star and triangle operators are defined as
\begin{equation}
    \vcenter{\hbox{\includegraphics[height=6.5em]{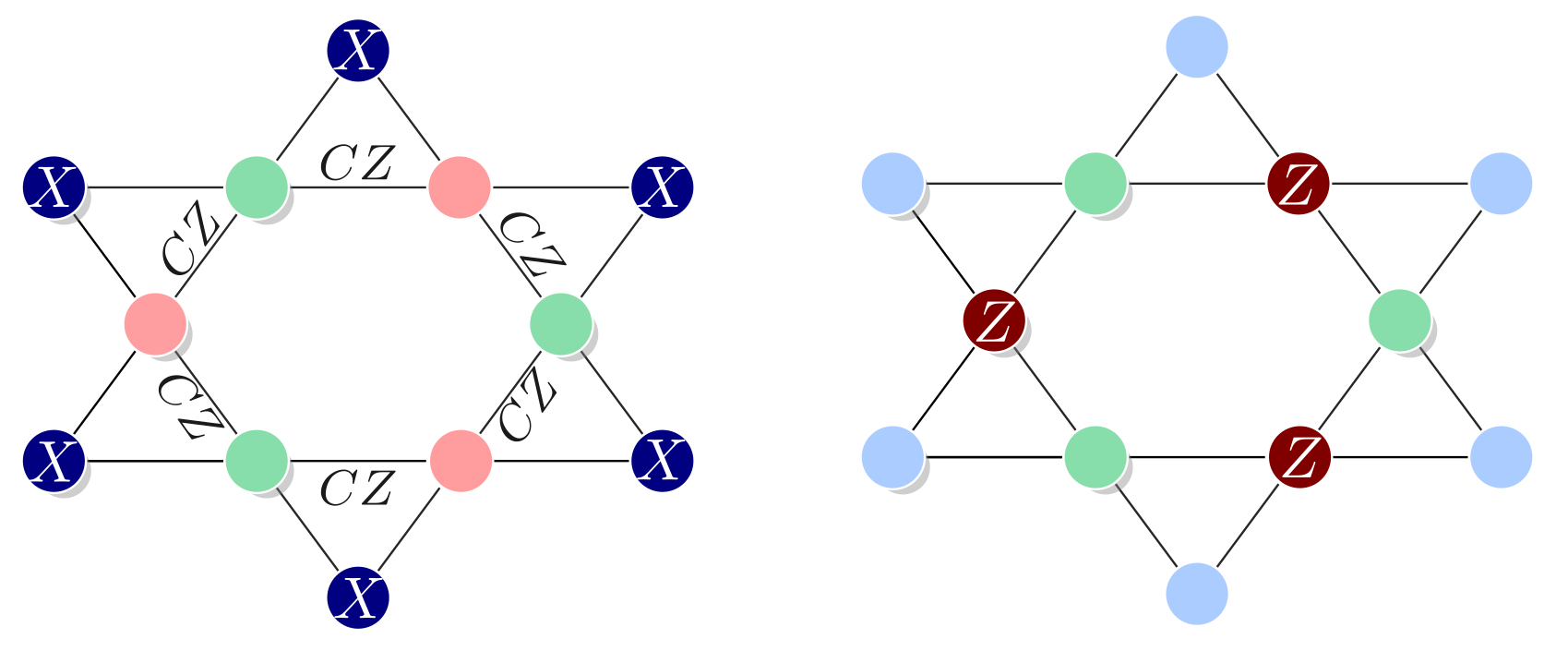}}}
    \label{Hamiltonian_terms}
\end{equation}
In this realization, the TO can be viewed as a $\mathbb Z_2^3$ gauge theory `twisted' to become non-Abelian \cite{Yoshida, Iqbal2024, D4preparation}. Each star and triangle term in the Hamiltonian has eigenvalues $\pm1$. Violations of the star operators create Abelian $e$-anyons, whereas violations of the triangle operators correspond to non-Abelian $m$-anyons. The anyons also carry color labels inherited from the lattice. Their fusion rules include $m_R\times m_R = 1 + e_B + e_G + e_B e_G$, $m_R\times e_B = m_R$, $m_R\times e_G = m_R$, $e_B\times e_B = 1$, $e_G\times e_G = 1$, together with their color permutations. In particular, the fusion of two non-Abelian anyons produces only Abelian fusion outcomes, making this TO an example of an acyclic anyon model \cite{Dauphinais2017}.\\
\indent Besides the existence of multiple fusion channels for the $m$-anyons, the non-Abelian nature of the TO is also reflected in the commutation relation
\begin{equation} \label{commutation}
    \vcenter{\hbox{\includegraphics[height=6.5em]{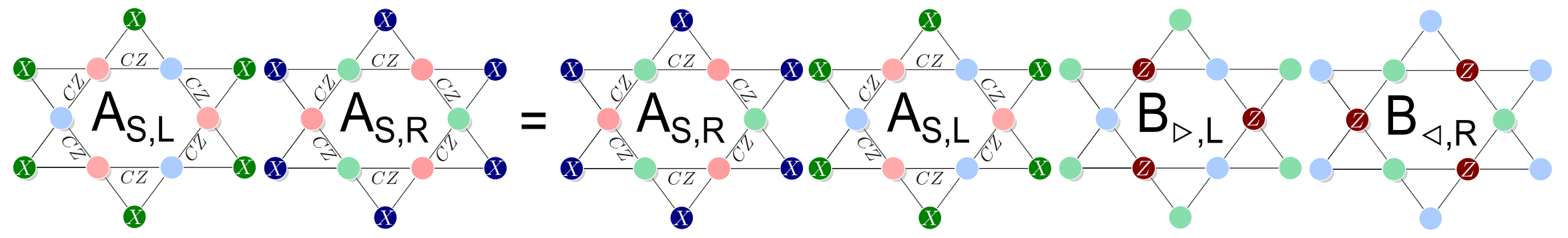}}}
\end{equation}
where $L$ and $R$ denote adjacent stars. Although all Hamiltonian terms commute in the absence of anyon excitations, the presence of an $m$-anyon causes the star operator whose support overlaps with the anyon to fail to commute with three neighboring star operators. Consequently, when measuring the anyon syndrome supplied as input to the ILP decoder, any star operator whose support contains an $m$-anyon should not be measured.

\subsection{Input}
\indent Let $S$ denote the set of sites of the syndrome lattice. In the realization of the $D_4$ TO on the three-colorable kagome lattice, these sites correspond to star locations, each of which supports one $e$-anyon of the same color as the star and two $m$-anyons of the other two colors. For each $m$-anyon color $a \in \{R,B,G\}$ and each $e$-anyon color $b \in \{R,B,G\}$, we are given:
\begin{itemize}
	\item Link sets: $L^{(m)}_a \subseteq \bigl\{\{r,s\}\mid r,s\in S,\ r\neq s\bigr\}$ and $L^{(e)}_b \subseteq \bigl\{\{r,s\}\mid r,s\in S,\ r\neq s\bigr\}$. A Pauli $\hat{X}$ error on a qubit of color $a$ pair-creates $m$-anyons of color $a$ along a link $\ell \in L^{(m)}_a$, while a Pauli $\hat{Z}$ error on a qubit of color $b$ pair-creates $e$-anyons of color $b$ along a link $\ell \in L^{(e)}_b$.
    \item Terminal sets: $T^{(m)}_a \subset S$ and $T^{(e)}_b \subset S$. These sets record the measured locations of all $m$-anyons and of the $e$-anyons on sites not occupied by an $m$-anyon. Unlike the lattice structure, the syndrome information varies across decoder instances.
	\item Link weights: $w^{(m)}_\ell > 0$ for $\ell \in L^{(m)}_a$ and $w^{(e)}_\ell > 0$ for $\ell \in L^{(e)}_b$.	These weights encode the local probabilities whose product gives $P(\bm{\sigma}|E)P(E)$. The values used in the ILP decoders for obtaining the error-correction phase diagram of the $D_4$ TO in Sec.~\ref{Sec:D4} are discussed in Appendix~\ref{App:D4decoder}.
\end{itemize}

\indent For later convenience, define the incident link sets $\partial^{(m)}_a s = \{ \ell \in L^{(m)}_a \mid s \in \ell \}$ and $\partial^{(e)}_b s = \{ \ell \in L^{(e)}_b \mid s \in \ell \}$ for each site $s \in S$. The links in $\partial^{(m)}_a s$ correspond to the three qubits of color $a$ in the interior of the star at site $s$, as shown in Eq.~\ref{Hamiltonian_terms}, on which a Pauli $\hat{X}$ operator creates an $m$-anyon of color $a$ at site $s$. Similarly, the links in $\partial^{(e)}_b s$ correspond to the six qubits of color $b$ at the tips of the star at site $s$, on which a Pauli $\hat{Z}$ operator creates an $e$-anyon of color $b$ at site $s$. We further express the syndromes using the binary terminal indicators
$$\tau^{(m)}_{a,s} =
\begin{cases}
	1, & s \in T^{(m)}_a, \\
	0, & \text{otherwise},
\end{cases}
\qquad
\tau^{(e)}_{b,s} =
\begin{cases}
	1, & s \in T^{(e)}_b, \\
	0, & \text{otherwise}.
\end{cases}$$

\subsection{Decision Variables}
\indent In this subsection, we list the decision variables of the ILP decoder, all of which are binary.\\
\paragraph{$m$-anyon variables.}
\begin{itemize}
	\item $y^{(m)}_{a,\ell}$: error-activation variable for $\ell \in L^{(m)}_a$. The value $y^{(m)}_{a,\ell}=1$ indicates that a Pauli $\hat{X}$ error pair-creates two $m$-anyons of color $a$ along link $\ell$, whereas $y^{(m)}_{a,\ell}=0$ indicates the absence of such an error.
	\item $x^{(m)}_{a,s}$: midpoint indicator for $m$-anyon strings at site $s$. The value $x^{(m)}_{a,s}=1$ indicates that two or three $m$-anyon strings of color $a$ are incident on site $s$, or equivalently, that the error-activation variables for two or three links $\ell \in \partial^{(m)}_a s$ equal 1, making $s$ a midpoint of an $m$-anyon string. In contrast, $x^{(m)}_{a,s}=0$ indicates that at most one $m$-anyon string of color $a$ is incident on $s$. On the three-colorable kagome lattice, no more than three $m$-anyon strings of the same color can be incident on any site, since $|\partial^{(m)}_a s|=3$. The midpoint indicator is therefore used primarily to enforce the correct number of incident $m$-anyon strings according to the $m$-anyon syndrome at site $s$.
\end{itemize}
\paragraph{$e$-anyon variables.}
\begin{itemize}
	\item $y^{(e)}_{b,\ell}$: error-activation variable for $\ell \in L^{(e)}_b$. The value $y^{(e)}_{b,\ell}=1$ indicates that a Pauli $\hat{Z}$ error pair-creates two $e$-anyons of color $b$ along link $\ell$, whereas $y^{(e)}_{b,\ell}=0$ indicates the absence of such an error.
	\item $x^{(e)}_{b,s}$: midpoint indicator for $e$-anyon strings at site $s$. The value $x^{(e)}_{b,s}=1$ indicates that two or three $e$-anyon strings of color $b$ are incident on site $s$, or equivalently, that the error-activation variables for two or three links $\ell \in \partial^{(e)}_b s$ equal 1, making $s$ a midpoint of an $e$-anyon string. In contrast, $x^{(e)}_{b,s}=0$ indicates that at most one $e$-anyon string of color $b$ is incident on $s$. On the three-colorable kagome lattice, up to six $e$-anyon strings of the same color can be incident on a site, since $|\partial^{(e)}_b s|=6$. However, two incident $e$-anyon strings generated by Pauli $\hat{Z}$ operators acting on adjacent tips of the star are equivalent, up to multiplication by a triangle operator $B_t$, to a single Pauli $\hat{Z}$ operator acting on the tip of a neighboring star. Since the link weights $w^{(e)}_\ell$ are positive for all $\ell \in L^{(e)}_b$, and the ILP decoder performs minimum-weight decoding, an optimal solution never contains activated error variables corresponding to adjacent tips of the same star. It is therefore sufficient to consider at most three incident $e$-anyon strings and to use a single binary midpoint indicator to enforce the correct number of incident $e$-anyon strings according to the $e$-anyon syndrome at site $s$.
    \item $T^{(e)}_{b,s}$: active $e$-anyon terminal indicator. The value $T^{(e)}_{b,s}=1$ indicates that a measured $e$-anyon of color $b$ at site $s$ is created solely by $e$-anyon strings, whereas $T^{(e)}_{b,s}=0$ indicates that either the measured $e$-anyon arises as a fusion outcome of non-Abelian $m$-anyons or no such $e$-anyon is measured at site $s$.
	\item $z^{(e)}_{b,s}$: inactive $e$-anyon terminal indicator. The value $z^{(e)}_{b,s}=1$ indicates that a measured $e$-anyon of color $b$ at site $s$ arises as a fusion outcome of non-Abelian $m$-anyons, whereas $z^{(e)}_{b,s}=0$ indicates that either the measured $e$-anyon is created solely by $e$-anyon strings or no such $e$-anyon is measured at site $s$.
\end{itemize}

\paragraph{Fusion indicators.} For each $e$-anyon terminal $s \in T^{(e)}_b$ and each $m$-anyon color $a$:
\begin{itemize}
	\item $g_{m_a\to e_b,s}$: fusion indicator. The value $g_{m_a\to e_b,s}=1$ indicates that a site $s$ containing a measured $e$-anyon of color $b$ is also a midpoint of an $m$-anyon string of color $a$. The measured $e$-anyon is therefore interpreted as a fusion outcome of non-Abelian $m$-anyons. This indicator is defined only for sites $s \in T^{(e)}_b$ and is used to determine the corresponding values of the $e$-anyon terminal indicators $T^{(e)}_{b,s}$ and $z^{(e)}_{b,s}$.
\end{itemize}

\subsection{Objective Function}
\indent For this ILP formulation, it is sufficient to assign weights only to the error-activation variables. In Appendix~\ref{App:D4decoder}, we discuss how these weights are chosen to account for both $P(\bm{\sigma}| E)$ and $P(E)$. All other decision variables therefore carry zero weight in the objective, which minimizes the total weight of the activated errors:
\begin{equation}
	\min
	\Bigg[
	\sum_{a=1}^3 \sum_{\ell \in L^{(m)}_a} w^{(m)}_\ell \, y^{(m)}_{a,\ell}
	+
	\sum_{b=1}^3 \sum_{\ell \in L^{(e)}_b} w^{(e)}_\ell \, y^{(e)}_{b,\ell}
	\Bigg].
\end{equation}

\subsection{Constraints}

\subsubsection{$m$-anyon String Constraints}

\indent For each $m$-anyon color $a$ and site $s$, the constraint enforces an even number of incident $m$-anyon strings of color $a$, either zero or two, when no $m$-anyon is measured at $s$, and an odd number, either one or three, when an $m$-anyon of color $a$ is measured at $s$, consistent with the acyclic property of the non-Abelian $m$-anyons. This permits open strings terminating at measured anyons, closed loops, and local three-way junctions:
\begin{equation}
\sum_{\ell \in \partial^{(m)}_a s} y^{(m)}_{a,\ell}
	=
	2 x^{(m)}_{a,s} + \tau^{(m)}_{a,s}.
	\label{eq:m-degree}
\end{equation}

\subsubsection{Fusion Constraints and $e$-Anyon Terminal Classification}

\indent The fusion indicators $g_{m_a\to e_b,s}$ are uniquely determined by the assignment of the error-activation variables $y^{(m)}_{a,\ell}$. At a site $s \in T^{(e)}_b$, the condition $g_{m_a\to e_b,s}=1$ requires that the measured $e$-anyon of color $b$ be located at a midpoint of an $m$-anyon string of color $a$, corresponding to an even number of incident $m$-anyon strings of color $a$. Since at most three $m$-anyon strings of the same color can be incident on a site of the kagome lattice, this number must be two. The decoder cannot assign either one or three incident $m$-anyon strings to a site with a measured $e$-anyon, because an odd number of incident strings would create an $m$-anyon at that site, whereas an $e$-anyon cannot be measured at the same site as an $m$-anyon. Such an error configuration would therefore be inconsistent with the measured $m$-anyon syndrome and is excluded by the constraints in Eq.~\ref{eq:m-degree}. Consequently, for each $e$-anyon terminal $s \in T^{(e)}_b$ and each $m$-anyon color $a$, it is sufficient to impose $g_{m_a\to e_b,s}=1$ whenever at least one error-activation variable associated with a link in $\partial^{(m)}_a s$ equals 1. This is enforced by
\begin{gather}
\sum_{\ell \in \partial^{(m)}_a s} y^{(m)}_{a,\ell}
	\le
	|\partial^{(m)}_a s|\; g_{m_a\to e_b,s}\qquad
	\sum_{\ell \in \partial^{(m)}_a s} y^{(m)}_{a,\ell}
	\ge
	g_{m_a\to e_b,s}
\end{gather}
\indent Whenever $g_{m_a\to e_b,s}=1$ for an $e$-anyon at site $s \in T^{(e)}_b$, the measured $e$-anyon is interpreted as a fusion outcome of non-Abelian $m$-anyons and is therefore designated as inactive by setting $z^{(e)}_{b,s}=1$. This is enforced by
\begin{align}
	z^{(e)}_{b,s} \leq \sum_{a=1}^3 g_{m_a\to e_b,s} \qquad
	z^{(e)}_{b,s} \geq g_{m_a\to e_b,s} \;\; \forall a
\end{align}
Otherwise, the measured $e$-anyon is active, with $T^{(e)}{b,s}=1$, indicating that it is created solely by $e$-anyon strings of the same color. Since a measured $e$-anyon must satisfy either $T^{(e)}{b,s}=1$ or $z^{(e)}_{b,s}=1$, the following constraint is imposed at every site $s$ and for each color $b$:
\begin{equation}
	T^{(e)}_{b,s} + z^{(e)}_{b,s} = \tau^{(e)}_{b,s}
\end{equation}
This constraint also ensures that both terminal indicators vanish at sites where no $e$-anyon of color $b$ is measured.

\subsubsection{$e$-Anyon String Constraints}
\indent Since the $e$-anyons are Abelian and pairwise annihilate to the vacuum, an odd number of incident $e$-anyon strings of color $b$ is required at a site $s$ where an $e$-anyon of the same color is measured, whereas an even number is required when no such $e$-anyon is measured. However, this parity condition does not apply at every site. If an odd number of $m$-anyon strings of any color is incident on $s$, then the site contains an $m$-anyon, and any number of $e$-anyon strings may also be incident because of the fusion rules $m_R\times e_B=m_R$ and $m_R\times e_G=m_R$, together with their color permutations. Furthermore, at a midpoint of $m$-anyon strings, where an even number of $m$-anyon strings of each color is incident, the local fusion state is a superposition of Abelian fusion outcomes according to $m_R\times m_R=1+e_B+e_G+e_Be_G$ and its color permutations. If such a midpoint corresponds to a star location of color $b$, the syndrome measurement yields either an $e$-anyon of color $b$ or the vacuum with equal probability, $\frac{1}{2}$. Incident $e$-anyon strings do not alter this superposition or the probability of measuring the $e$-anyon. Therefore, the following constraint is imposed only at sites $s$ with no incident $m$-anyon strings:
\begin{equation}
\sum_{\ell \in \partial^{(e)}_b s} y^{(e)}_{b,\ell}
	=
	2 x^{(e)}_{b,s} + T^{(e)}_{b,s}
	\label{eq:e-ideal}
\end{equation}
At sites with incident $m$-anyon strings, the $e$-anyon strings are instead allowed to terminate or branch freely.\\
\indent Whether an $m$-anyon string is incident on a site $s$ depends on the values of the error-activation variables $y^{(m)}_{a,\ell}$. Since the inclusion of an ILP constraint cannot itself depend on decision-variable assignments, Eq.~\ref{eq:e-ideal} is implemented in practice using a big-$M$ relaxation.\\
\indent The big-$M$ relaxation is implemented by first defining
\[
S^{(m)}_s = \sum_{a \in \{R,B,G\}} x^{(m)}_{a,s},
\qquad
T^{(m)}_s = \sum_{a \in \{R,B,G\}} \tau^{(m)}_{a,s}.
\]
Here, $S^{(m)}_s >0$ indicates the presence of an $m$-anyon string midpoint at site $s$, while $T^{(m)}_s >0$ indicates the presence of a measured $m$-anyon. Then, for each $e$-anyon color $b$ and site $s$, we choose $M\geq \max_s|\partial^{(e)}_b s|=6$ and impose
\begin{align}
\sum_{\ell \in \partial^{(e)}_b s} y^{(e)}_{b,\ell}
	- 2 x^{(e)}_{b,s}
	- T^{(e)}_{b,s}
	- M S^{(m)}_s
	&\leq M T^{(m)}_s,\\
    -\sum_{\ell \in \partial^{(e)}_b s} y^{(e)}_{b,\ell}
	+ 2 x^{(e)}_{b,s}
	+ T^{(e)}_{b,s}
	- M S^{(m)}_s
	&\leq
	M T^{(m)}_s.
\end{align}
\indent When a site $s$ has neither a measured $m$-anyon nor an $m$-anyon string midpoint, $S^{(m)}_s=T^{(m)}_s=0$, and these constraints reduce exactly to Eq.~\ref{eq:e-ideal}. Otherwise, $S^{(m)}_s+T^{(m)}_s\geq1$, and the constraints are relaxed so that they impose no restriction on the $e$-anyon degree. Indeed, they are then automatically satisfied because
\[
\sum_{\ell \in \partial^{(e)}_b s} y^{(e)}_{b,\ell} - 2 x^{(e)}_{b,s} - T^{(e)}_{b,s} \leq \sum_{\ell \in \partial^{(e)}_b s} y^{(e)}_{b,\ell} \leq \max_s|\partial^{(e)}_b s| \leq M \leq M S^{(m)}_s + M T^{(m)}_s
\]

\section{MWPM and ILP Decoding Protocols for the $D_4$ Topological Order}
\label{App:D4decoder}
\indent In this appendix, we provide the detailed protocols for the MWPM decoder and the three ILP decoders used to determine the error-correction threshold of the $D_4$ TO with perfect anyon syndromes in Sec.~\ref{Sec:D4} of the main text. Since the $m$-anyons are acyclic non-Abelian anyons, they can be removed in a single correction step by pairwise annihilation. This process may leave behind Abelian $e$-anyons as fusion outcomes, which require a second round of pairwise annihilation. Therefore, all error-correction protocols considered for the $D_4$ TO with perfect anyon syndromes consist of two steps.\\
\indent Simulations of the error-correction protocol, including error introduction, syndrome measurement, and correction, are performed using the stabilizer tableau formalism developed in Ref.~\cite{chirame2024stabilizingnonabeliantopologicalorder}. A logical error is declared if, at the end of the protocol, a logical stabilizer of the initial state has either flipped sign or is no longer a stabilizer.

\subsection{MWPM and ILP Decoding Protocols}
\subsubsection{MWPM Decoder}
\indent After a single-qubit Pauli channel with $p_y=p_xp_z$ independently generates all three colors of $m$- and $e$-anyons on the initial state, MWPM decoding of the $D_4$ TO proceeds as follows:
\begin{itemize}[label={}]
    \item 1. Measure all colors of $m$-anyons.
    \item 2. Perform MWPM separately for each $m$-anyon color, assigning uniform weight to all links. The decoder outputs a set of links on which Pauli $\hat{X}$ operators are applied to pairwise annihilate the $m$-anyons.
    \item 3. Measure all colors of $e$-anyons.
    \item 4. Perform MWPM separately for each $e$-anyon color. The decoder outputs a set of links on which Pauli $\hat{Z}$ operators are applied to pairwise annihilate the $e$-anyons. Links connecting two $e$-anyon locations along an $m$-anyon correction path are assigned zero weight, while all other links are assigned uniform weight.
\end{itemize}

\indent On the kagome lattice, the weight assignment in Step~4 is implemented as follows. Whenever the $m$-anyon correction in Step~2 applies Pauli $\hat{X}$ operators to both qubits of the same color (blue) on one side of an hourglass, the $e$-anyon link corresponding to the central (green) qubit of that hourglass is assigned zero weight:
\begin{equation} \label{hourglass}
    \vcenter{\hbox{\includegraphics[height=6em]{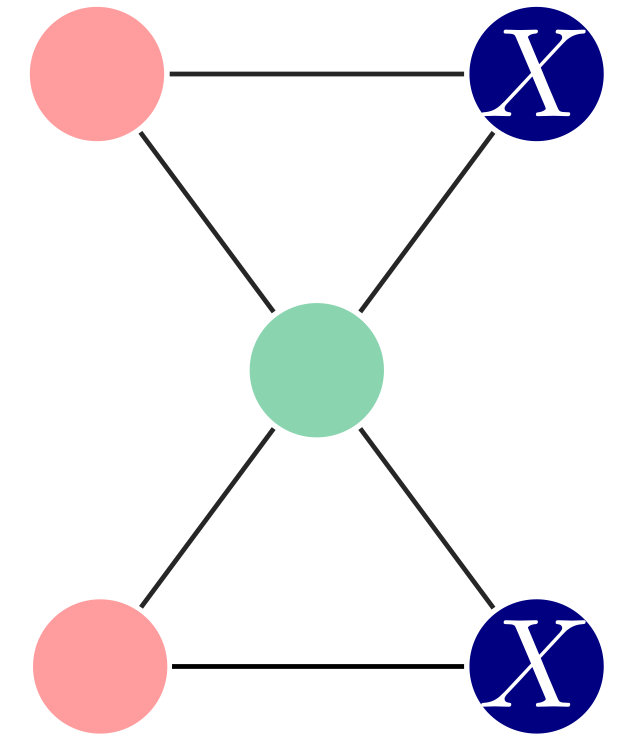}}}
\end{equation}
Because the correction of non-Abelian $m$-anyons in Step~2 is implemented by finite-depth Pauli $\hat{X}$ strings, it can create additional $e$-anyons along the correction paths. The $e$-anyons should therefore be pairwise annihilated preferentially along the same paths. This zero-weight assignment is thus adopted to allow the $e$-anyons to propagate freely along the $m$-anyon correction paths. Since the same consideration applies to any two-step error-correction scheme for the $D_4$ TO, we use the same weight assignment for all three ILP decoders described below. In practice, this choice improves the error-correction threshold of both the MWPM and ILP decoders.

\subsubsection{ILP Decoder for Maximizing $P(E)$}
\indent After a single-qubit Pauli channel with $p_y=p_xp_z$ independently generates all three colors of $m$- and $e$-anyons on the initial state, ILP decoding that maximizes $P(E)$ proceeds as follows:
\begin{itemize}[label={}]
    \item 1. Measure all colors of $m$-anyons. Then, measure all $e$-anyons at sites where no $m$-anyon is measured.
    \item 2. Perform ILP decoding using the formulation in Appendix~\ref{App:D4ILP}, with $w^{(m)}_\ell = \ln \frac{1-p_x}{p_x}$ and $w^{(e)}_\ell = \ln \frac{1-p_z}{p_z}$. This weight assignment accounts only for $P(E)$. The decoder outputs a set of links with $y^{(m)}_{a,\ell}=1$ and $y^{(e)}_{b,\ell}=1$. However, only Pauli $\hat{X}$ operators are applied to the qubits corresponding to links with $y^{(m)}_{a,\ell}=1$; no correction is applied to those corresponding to links with $y^{(e)}_{b,\ell}=1$.
    \item 3. Measure all colors of $e$-anyons.
    \item 4. Perform MWPM separately for each $e$-anyon color. The decoder outputs a set of links on which Pauli $\hat{Z}$ operators are applied to pairwise annihilate the $e$-anyons. Links connecting two $e$-anyon locations along an $m$-anyon correction path are assigned zero weight, while all other links are assigned uniform weight.
\end{itemize}

\indent The protocol described above yields a higher error-correction threshold than the variant in which Step~2 also applies Pauli $\hat{Z}$ operators to the links with $y^{(e)}_{b,\ell}=1$, in addition to applying Pauli $\hat{X}$ operators to the links with $y^{(m)}_{a,\ell}=1$. This behavior is also observed for the other two ILP decoders described below and is justified in the next subsection.

\subsubsection{ILP Decoder for Maximizing $P(E|\sigma)$}
\indent After a single-qubit Pauli channel with $p_y=p_xp_z$ independently generates all three colors of $m$- and $e$-anyons on the initial state, ILP decoding that maximizes $P(E|\bm{\sigma})$ proceeds as follows:
\begin{itemize}[label={}]
    \item 1. Measure all colors of $m$-anyons. Then, measure all $e$-anyons at sites where no $m$-anyon is measured.
    \item 2. Perform ILP decoding using the formulation in Appendix~\ref{App:D4ILP} with $w^{(e)}_\ell = \ln \frac{1-p_z}{p_z}$. The $m$-anyon link weights depend on the measured non-Abelian $m$-anyon syndromes at the two sites connected by $\ell$. If both sites contain an $m$-anyon of any color, assign $w^{(m)}_\ell=\ln\frac{1-p_x}{p_x}$. If exactly one of the two sites contains an $m$-anyon of any color, assign $w^{(m)}_\ell=\ln\frac{\sqrt{2}(1-p_x)}{p_x}$. If neither site contains an $m$-anyon, assign $w^{(m)}_\ell=\ln\frac{2(1-p_x)}{p_x}$. The decoder outputs a set of links with $y^{(m)}_{a,\ell}=1$ and $y^{(e)}_{b,\ell}=1$. Pauli $\hat{X}$ operators are applied only to the qubits corresponding to links with $y^{(m)}_{a,\ell}=1$, while no correction is applied to those corresponding to links with $y^{(e)}_{b,\ell}=1$.
    \item 3. Measure all colors of $e$-anyons.
    \item 4. Perform MWPM separately for each $e$-anyon color. The decoder outputs a set of links on which Pauli $\hat{Z}$ operators are applied to pairwise annihilate the $e$-anyons. Links connecting two $e$-anyon locations along an $m$-anyon correction path are assigned zero weight, while all other links are assigned uniform weight.
\end{itemize}

\indent At each site along a string of Pauli $\hat{X}$ errors, $m$-anyons fuse into a superposition of the vacuum and an $e$-anyon. Measuring the $e$-anyon syndrome at such a site collapses this superposition, yielding either the vacuum or an $e$-anyon with equal probability $p_{f\to a}=\frac{1}{2}$, where $a$ denotes either outcome. Therefore, requiring the intermediate fusion outcomes along an $m$-anyon string to be consistent with the measured syndromes contributes an additive factor of $\ln 2$ to the objective for each site along the string that does not contain an $m$-anyon and at which an $e$-anyon measurement is therefore performed. Equivalently, since each midpoint of an $m$-anyon string has exactly two incident $m$-anyon links of the corresponding color, this factor can be split between the two links, giving a penalty of $\ln\sqrt{2}$ to each link-activation variable $y^{(m)}_{a,\ell}$. This reproduces the weight assignment in Step~2 of the protocol described above. In this way, weights are assigned only to link-activation variables while still accounting for $P(E|\bm{\sigma})\propto P(\bm{\sigma}|E)P(E)$.\\
\indent If the error creates an isolated, homologically trivial, closed $m$-anyon loop, no $m$-anyons can be measured on the loop because of the acyclic fusion structure. However, requiring such an isolated component to fuse to the vacuum imposes an additional global consistency condition in which the parity of the measured $e$-anyons of each color on the loop must be even. Admittedly, this condition is not incorporated into the weight assignment in Step~2. However, a related study \cite{jing2025intrinsicheralding} found that such isolated, homologically trivial, closed $m$-anyon loops occur only rarely, and neglecting them has a minimal effect on the numerical error-correction threshold.

\subsubsection{ILP Decoder with Effective Weight Ratio}
\indent After a single-qubit Pauli channel with $p_y=p_xp_z$ independently generates all three colors of $m$- and $e$-anyons on the initial state, ILP decoding with effective weight ratio proceeds as follows:
\begin{itemize}[label={}]
    \item 1. Measure all colors of $m$-anyons. Then, measure all $e$-anyons at sites where no $m$-anyon is measured.
    \item 2. Perform ILP decoding using the formulation in Appendix~\ref{App:D4ILP}, with a constant effective weight ratio $r_{\mathrm{eff}}=w^{(m)}_\ell/ w^{(e)}_\ell$. The optimal value of $r_{\mathrm{eff}}$ is determined by trial and error. The decoder outputs a set of links with $y^{(m)}_{a,\ell}=1$ and $y^{(e)}_{b,\ell}=1$. Pauli $\hat{X}$ operators are applied only to the qubits corresponding to links with $y^{(m)}_{a,\ell}=1$, while no correction is applied to those corresponding to links with $y^{(e)}_{b,\ell}=1$.
    \item 3. Measure all colors of $e$-anyons.
    \item 4. Perform MWPM separately for each $e$-anyon color. The decoder outputs a set of links on which Pauli $\hat{Z}$ operators are applied to pairwise annihilate the $e$-anyons. Links connecting two $e$-anyon locations along an $m$-anyon correction path are assigned zero weight, while all other links are assigned uniform weight.
\end{itemize}

\subsection{Deferred Correction of $e$-Anyons}
\indent In this subsection, we justify why the $D_4$ QEC threshold is higher when only Pauli $\hat{X}$ operators are applied in Step~2 of the ILP protocols to correct the $m$-anyons, while all correction of the $e$-anyons is deferred to Step~4. The relevant quantum channels in the ILP decoding protocols are:
\begin{itemize}
    \item the single-qubit Pauli channel describing physical noise, $\mathcal{P}$;
    \item the $m$- and $e$-anyon syndrome measurements in Step~1, $\mathcal{M}_{ILP}$;
    \item the application of Pauli $\hat{X}$ operators to correct the $m$-anyons in Step~2, $\mathcal{E}_{m,ILP}$;
    \item the optional application of Pauli $\hat{Z}$ operators to correct the $e$-anyons in Step~2, $\mathcal{E}_{e,ILP}$;
    \item the $e$-anyon syndrome measurement in Step~3, $\mathcal{M}_{e}$.
\end{itemize}

\indent Whether Pauli $\hat{Z}$ operators are applied in Step~2 does not affect the correction of the $m$-anyons by Pauli $\hat{X}$ operators. Indeed, our numerical simulations show that the threshold improvement obtained by deferring $e$-anyon correction to Step~4 arises entirely from suppressing the proliferation of $e$-anyons, while the proliferation of $m$-anyons remains unchanged. It therefore suffices to consider only the $e$-anyon threshold in the argument below.\\
\indent If the correction of $e$-anyons is deferred from Step~2 to Step~4, the density matrix immediately before Step~4 is
\begin{equation}
    \mathcal{E}_{e_{\mathrm{eff}}1}\left(\hat{\rho}\right) = \mathcal{M}_{e} \circ \mathcal{E}_{m,ILP} \circ \mathcal{M}_{ILP} \circ \mathcal{P} \left(\hat{\rho}\right),
\end{equation}
where $\hat{\rho}$ denotes the density matrix of the initial state. Optimal correction of the $e$-anyons then selects the homology class with the largest diagonal weight in this density matrix.\\
\indent In contrast, if $e$-anyon correction is applied in Step~2, the density matrix immediately before Step~4 is
\begin{equation}
    \mathcal{E}_{e_{\mathrm{eff}}2}\left(\hat{\rho}\right) = \mathcal{M}_{e} \circ \mathcal{E}_{e,ILP} \circ \mathcal{E}_{m,ILP} \circ \mathcal{M}_{ILP} \circ \mathcal{P} \left(\hat{\rho}\right).
\end{equation}
The order of $\mathcal{E}_{e,ILP}$ and $\mathcal{E}_{m,ILP}$ is arbitrary because the two channels commute, $\mathcal{E}_{e,ILP} \circ \mathcal{E}_{m,ILP} = \mathcal{E}_{m,ILP} \circ \mathcal{E}_{e,ILP}$.\\
\indent Furthermore, the $e$-anyon correction channel $\mathcal{E}_{e,ILP}$ commutes with the $e$-anyon syndrome measurement channel $\mathcal{M}_{e}$. Measuring the $e$-anyon syndromes first, thereby collapsing coherence between $e$-anyons and the vacuum, and then applying Pauli $\hat{Z}$ operators yields the same final state as applying the Pauli $\hat{Z}$ operators before the measurement. Hence,
\begin{equation}
    \mathcal{E}_{e_{\mathrm{eff}}2}\left(\hat{\rho}\right) = \mathcal{E}_{e,ILP} \circ \mathcal{M}_{e} \circ  \mathcal{E}_{m,ILP} \circ \mathcal{M}_{ILP} \circ \mathcal{P} \left(\hat{\rho}\right) = \mathcal{E}_{e,ILP} \circ \mathcal{E}_{e_{\mathrm{eff}}1}\left(\hat{\rho}\right).
\end{equation}
\indent Optimal decoding based on $\mathcal{E}_{e_{\mathrm{eff}}2}\left(\hat{\rho}\right)$ selects the $e$-anyon homology class with the largest diagonal element. Viewed instead as a decoder acting on $\mathcal{E}_{e_{\mathrm{eff}}1}\left(\hat{\rho}\right)$, this procedure is generally non-optimal, since the additional channel $\mathcal{E}_{e,ILP}$ can alter the relative weights of the homology classes. Its threshold therefore cannot exceed that of the optimal decoder acting directly on $\mathcal{E}_{e_{\mathrm{eff}}1}\left(\hat{\rho}\right)$.\\
\indent It is reasonable to expect the same argument to apply to minimum-weight decoding implemented by the ILP decoders. Correcting the $e$-anyons in Step~2 amounts to performing minimum-weight decoding based on $\mathcal{E}_{e_{\mathrm{eff}}2}\left(\hat{\rho}\right)$, which is expected to have a lower threshold than minimum-weight decoding performed directly on $\mathcal{E}_{e_{\mathrm{eff}}1}\left(\hat{\rho}\right)$, for which the optional $e$-anyon correction channel $\mathcal{E}_{e,ILP}$ is absent.
\end{document}